\documentclass[twocolumn]{autart}    
\usepackage{graphicx}      
\usepackage{amsmath,amsfonts,amsxtra,amssymb,mathtools}
\usepackage{xparse,etoolbox}
\usepackage{pbox}
\usepackage{pgfplots}
\usepackage{multirow}
\usepackage{xspace,xcolor}
\usepackage{subcaption}
\usepackage{siunitx}
\usepackage{booktabs,tabularx}
\usepackage{datatool}
\usepackage{calc}

\DTLsetseparator{ =}
\newcommand{\DTLfetchsave}[5]{%
	\edtlgetrowforvalue{#2}{\dtlcolumnindex{#2}{#3}}{#4}%
	\dtlgetentryfromcurrentrow{\dtlcurrentvalue}{\dtlcolumnindex{#2}{#5}}%
	\let#1\dtlcurrentvalue
}

\renewcommand{\d}{\mathrm{d}}

\newcommand{\dzeta}{\partial_\zeta}
\newcommand{\dz}{\partial_z}
\newcommand{\dt}{\partial_t}
\newcommand{\dtau}{\partial_\tau}

\newcommand{\adj}{\mathop{\mathrm{adj}}}

\NewDocumentCommand{\R}{O{}}{{\mathbb{R}^{#1}}}	

\newcommand{\transpose}{\top}				
\NewDocumentCommand{\MATRIX}{s m}
{%
	\IfBooleanTF{#1}{
	\begin{bsmallmatrix}#2\end{bsmallmatrix}		
	}{
	\begin{bmatrix}#2\end{bmatrix}
	}
}

\NewDocumentCommand{\INT}{s m m m m}
{%
	\IfBooleanT{#1}{\textstyle}\int_{#2}^{#3} \! #4 \mathrm{d}#5
}		
\newcommand{\rank}{\textrm{rank}\,}				
\NewDocumentCommand{\diag}{s m}{
	\IfBooleanTF{#1}{\textrm{diag}(#2)}{\textrm{diag}\left(#2\right)}
}

\NewDocumentCommand{\abs}{m}{
	\lvert\ifblank{#1}{\:\cdot\:}{#1}\rvert
	}											

\NewDocumentCommand{\var}{s m O{} m o}{
	#2^{\mathrm{#3}}_{\mathrm{#4}\IfValueTF{#5}{\ifblank{#4}{}{,}#5}{}}
	\IfBooleanTF{#1}{}{(z)}
	}%
\NewDocumentCommand \vect { s m }{
	\lbrack\vectentry[\enspace]{#2}
	\rbrack\IfBooleanT{#1}{^\transpose}
}%
\NewDocumentCommand{\col}{s m}{
	\mathrm{col}%
	\IfBooleanTF {#1}{%
		(\vectentry[$, $]{#2}[])%
	}{%
	\left(\vectentry[,\,]{#2}[]\right)
	}
}%
\NewDocumentCommand{\domain}{s m}{
	\IfBooleanTF {#1}{(}{\lbrack}
	\vectentry[,]{#2}[] 	
	\IfBooleanTF {#1}{)}{\rbrack}
	}%
\NewDocumentCommand{\innerp}{m}{
	\langle\vectentry[,]{#1}[]\rangle
}			
\ExplSyntaxOn									
\NewDocumentCommand{\vectentry}{O{\enspace} m O{\,}}{
	#3 \dbacc_values:n {#1}{#2} #3
	}%
\cs_new_protected:Npn \dbacc_values:n #1 #2
{
	\seq_set_split:Nnn \l_tmpa_seq { & } {#2}	
	\seq_use:Nn \l_tmpa_seq {#1}
}
\ExplSyntaxOff

\newcommand{\wrt}{w.\,r.\,t.\xspace}			

\renewcommand{\tt}{(t)}

\NewDocumentCommand{\Time}{s}{\mathrm{I}\IfBooleanTF{#1}{^*}{}}
\DTLloaddb[noheader, keys={key,value}]{dataTable}{data/parameters.dat}
\NewDocumentCommand{\val}{s o m}{%
	\DTLfetchsave{\temp}{dataTable}{key}{#3}{value}\IfBooleanTF{#1}{	\IfNoValueTF{#2}{\num{\temp}}{\SI{\temp}{#2}}}{\temp}%
}

\newcommand{\laplace}{\operatorname{\mbox{\setlength{\unitlength}{0.1em}%
		\begin{picture}(20,10)%
		\put(2,3){\circle{4}}%
		\put(4,3){\line(1,0){13}}%
		\put(18,3){\circle*{4}}%
		\end{picture}%
	}}%
}%
\newcommand{\Laplace}{\operatorname{\mbox{\setlength{\unitlength}{0.1em}%
		\begin{picture}(20,10)%
		\put(2,3){\circle*{4}}%
		\put(3,3){\line(1,0){13}}%
		\put(18,3){\circle{4}}%
		\end{picture}%
	}}%
}%

\pgfplotsset{%
	compat=1.5,
	every axis plot/.append style={thick},
	/tikz/font=\footnotesize,
	invoke before crossref tikzpicture={\tikzexternaldisable},
	invoke after crossref tikzpicture={\tikzexternalenable},
	/pgfplots/enlarge x limits=false,
}
\usetikzlibrary{external,
	calc,
	backgrounds,
	patterns,
	arrows,
	matrix,
	intersections,
	positioning,
	fit,
	arrows.meta,
	shapes.geometric,
	spy}

\usetikzlibrary{spy}

\tikzset{external/up to date check=md5}
\DeclareRobustCommand{\plotref}[1]{\tikzexternaldisable(\ref{#1})\tikzexternalenable} %
\AtBeginDocument{\tikzexternaldisable}
\NewDocumentEnvironment{tikzexternal}{s o O{false}}{%
	\IfNoValueF{#1}{\tikzsetnextfilename{#2}}
	\tikzset{external/remake next=#3}
	\tikzexternalenable
	\IfBooleanTF{#1}{\vspace{1em plus 1em}}{\vspace{0em plus .5em minus .5em}}
}{%
	\vspace{0em minus .5em}
	\tikzexternaldisable
}

\pgfplotsset{every axis/.style={
			width=8cm,
			height=4cm,
			xlabel={$t$},
			xmajorgrids,
			ymajorgrids,
			clip=false,
		}}

\tikzset{external/force remake=false}

\NewDocumentCommand \markDomain { s m m O{bgColor}}{%
	\begin{pgfonlayer}{background}
	\fill[#4, fill opacity=0.2]
	({rel axis cs:0,0}-|{axis cs:#3,0})
	rectangle
	({rel axis cs:0,1}-|{axis cs:#2,0});
	\end{pgfonlayer}
}%

\NewDocumentCommand{\noIdentificationAreaLabel}{O{bgColor}}{%
\tikzexternaldisable\tikz[]{\node[draw, outer sep = 0, fill = #1, fill opacity=0.2]{}}\tikzexternalenable\xspace
}

\DTLfetchsave{\fT}{dataTable}{key}{T}{value}
\DTLfetchsave{\foI}{dataTable}{key}{occurrence1}{value}
\DTLfetchsave{\foIT}{dataTable}{key}{occurrence1T}{value}		
\DTLfetchsave{\foII}{dataTable}{key}{occurrence2}{value}
\DTLfetchsave{\foIIT}{dataTable}{key}{occurrence2T}{value}
\DTLfetchsave{\foIII}{dataTable}{key}{occurrence3}{value}
\DTLfetchsave{\foIIIT}{dataTable}{key}{occurrence3T}{value}
\newcommand{\markDomains}{		
	\markDomain{0}{\fT}
	\markDomain{\foI}{\foIT}
	\markDomain{\foII}{\foIIT}
	\markDomain{\foIII}{\foIIIT}}

\DTLfetchsave{\tDI}{dataTable}{key}{tDetection1}{value}
\DTLfetchsave{\tDII}{dataTable}{key}{tDetection2}{value}
\DTLfetchsave{\tDIII}{dataTable}{key}{tDetection3}{value}
\DTLfetchsave{\tDIT}{dataTable}{key}{tDetection1T}{value}
\DTLfetchsave{\tDIIT}{dataTable}{key}{tDetection2T}{value}
\DTLfetchsave{\tDIIIT}{dataTable}{key}{tDetection3T}{value}
\DTLfetchsave{\tEnd}{dataTable}{key}{tEnd}{value}

\DTLfetchsave{\tTDI}{dataTable}{key}{tDetectionT1}{value}
\DTLfetchsave{\tTDII}{dataTable}{key}{tDetectionT2}{value}
\DTLfetchsave{\tTDIII}{dataTable}{key}{tDetectionT3}{value}

\colorlet{RED}{red!90!green}
\colorlet{GREEN}{red!25!green}
\colorlet{BLUE}{blue}
\colorlet{VIOLET}{violet}
\colorlet{bgColor}{blue!50}

\newcommand{\red}[1]{\textcolor{RED}{#1}}

\usepackage{fixme}
\usepackage[normalem]{ulem}
\fxsetup{
	status=draft,
	author=,
	layout=footnote, 
	theme=color,
}
\definecolor{fxnote}{named}{red}

\renewcommand*{\thefootnote}{\textcolor{fxnote}{\alph{footnote}}}

\newcommand\redout{\bgroup\markoverwith
	{\textcolor{red}{\rule[.5ex]{2pt}{0.4pt}}}\ULon}

\usepackage{adjustbox}
\usepackage{nicefrac}
\usepackage{mathdots}
\usepackage[subnum]{cases}
\selectcolormodel{cmyk}

\newcommand{\n}{-}
\newcommand{\p}{+}

\renewcommand{\exp}[1]{\mathrm{e}^{#1}}

\newcommand{\z}{\bar{z}}

\begin{document}
\begin{filecontents*}{data/parameters.dat}
approxError = 0.0001
successizeSteps = 12
backsteppingGrid = 51
dt = 0.003
"T" = 30
tEnd = 279.999
identification_dt = 0.003
detection_dt = 0.003
T0 = 5.5
tau_plus = 2.75
tau_mins = -2.75
T_tau_min = 27.25
T_tau_plus = 32.75
bound1 = 0.7
bound2 = 0.3
fb1 = 4.20209684045367
fb2 = 8.10232275502489
fb3 = 11.7918101186306
occurrence1 = 70
occurrence1T = 100
occurrence2 = 140
occurrence2T = 170
occurrence3 = 210
occurrence3T = 240
tDetection1 = 75.516
tDetection1T = 105.516
tDetection2 = 188.937
tDetection2T = 218.937
tDetection3 = 211.323
tDetection3T = 241.323
tDetectionT1 = 100.002
tDetectionT2 = 188.937
tDetectionT3 = 240
\end{filecontents*}

\begin{frontmatter}
\runtitle{Fault diagnosis for heterodirectional
hyperbolic systems}  

\title{Fault diagnosis for linear heterodirectional
hyperbolic ODE-PDE systems using backstepping-based trajectory planning}


\author{Ferdinand Fischer\corauthref{corres}}\ead{ferdinand.fischer@uni-ulm.de}, 
\author{Joachim Deutscher}\ead{joachim.deutscher@uni-ulm.de}

\corauth[corres]{Corresponding author}
	
\address{Institut f\"ur Mess-, Regel- und Mikrotechnik, Universit\"at Ulm, Albert-Einstein-Allee 41, D-89081 Ulm, Germany}%
%

\begin{abstract}                          
This paper is concerned with the fault diagnosis problem for general linear heterodirectional hyperbolic ODE-PDE systems. A systematic solution is presented for additive time-varying actuator, process and sensor faults in the presence of disturbances. The faults and disturbances are represented by the solutions of finite-dimensional signal models, which allow to take a large class of signals into account. For disturbances, that are only bounded, a threshold for secured fault diagnosis is derived. By applying integral transformations to the system an algebraic fault detection equation to detect faults in finite time is obtained. The corresponding integral kernels result from the realization of a finite-time transition between a non-equilibrium initial state and a vanishing final state of a hyperbolic ODE-PDE system. For this new challenging problem, a systematic trajectory planning approach is presented. In particular, this problem is facilitated by mapping the kernel equations into backstepping coordinates and tracing the solution of the transition problem back to a simple trajectory planning. The fault diagnosis for a $4\times 4$ heterodirectional hyperbolic system coupled with a second order ODE demonstrates the results of the paper.
\end{abstract}

\end{frontmatter}

\section{Introduction}
\subsection{Background and motivation}
General linear heterodirectional hyperbolic systems consist of transport PDEs propagating in both the negative and positive direction of the spatial coordinate. If, in addition, a finite-dimensional system is coupled to the PDE system, then a so-called ODE-PDE system results. This type of systems appears in many applications. Examples are coupled string networks (see, e.g., \cite[Ch. 6]{Luo1999}) and  networks of open channels and transmission lines (see, e.g., \cite{Bastin2016}) with dynamic boundary conditions. 
Increasing demands on the achievable control performance require the application of more and more advanced control methods (see, e.g., \cite{Bastin2016} for an overview). 
To ensure a safe operation of these resulting control systems, the fault diagnosis is crucial. As far as distributed-parameters systems (DPS) are considered, available methods mainly focus on observer-based fault diagnosis (see \cite{Ghantasala2009,Feng2019} for the early-lumping and \cite{Demetriou2002,Dey2019,Xu2019} for the late lumping approach). These approaches require at least for the implementation an approximation of the obtained observer. Consequently, a high observer order is required to avoid spillover effects for a reliable fault detection. The few alternatives to these approaches are the fault diagnosis based on a functional observer described in \cite{Deutscher2016} or the algebraic approach in \cite{Gehring2016}. These methods, suffer from a difficult design or the fact that disturbances are not taken into account (see \cite{Fischer2020} for a more detailed discussion). Recently, a new algebraic fault diagnosis method was presented in \cite{Fischer2020}. In particular, an explicit expression is derived for the faults, which require only known signals. With this, it is possible to determine the faults in prescribed finite-time. Both deterministic disturbances as well as disturbances with a known bound can easily be incorporated in the fault detection. Another advantage of the approach is that it can be efficiently implemented by making use of finite impulse response filters. So far, parabolic and biharmonic PDE-systems were considered in \cite{Fischer2020} while results for hyperbolic systems are limited so far to wave equations in \cite{Fischer2020a}. However, these results do neither directly extend to general linear heterodirectional hyperbolic systems nor to ODE-PDE systems.

\subsection{Contribution}
In this paper, a solution for the fault diagnosis problem, i.e., fault detection, isolation and identification, for general linear heterodirectional hyperbolic ODE-PDE systems is presented. The unknown faults can be additive actuator, sensor or process faults. In addition, unknown but bounded disturbances can be present. Different from the previous result \cite{Fischer2020}, both the faults and the disturbances are assumed to be solutions of a finite-dimensional linear time-invariant signal model. This allows to take a significant larger class of signal types for faults and disturbances into account, increasing the flexibility to the fault diagnosis scheme. More precisely, polynomial and trigonometric signals or combinations thereof can be easily considered.
Furthermore, the design of the explicit expressions for the faults become much simpler.
The new approach extends the results in \cite{Fischer2020} to multiple faults with different signal types and fewer restrictions to the number of available measurements. The fault identification is achieved in finite time by means of simple to implement expressions without the approximation of a DPS. 
Under the influence of a bounded, but otherwise arbitrary disturbance, the fault detection, isolation and estimation is achieved by introducing a threshold. 

The presented approach is based on the method introduced in \cite{Fischer2020}. An algebraic input-output expression is used for the fault diagnosis, which follows from the application of integral transformations. The kernels of these transformations are chosen so that the considered fault diagnosis problem can be solved. This leads to so-called fault diagnosis kernel equations in the form of a general linear heterodirectional hyperbolic ODE-PDE system with given initial and end conditions as well as an input as degree of freedom. Consequently, the kernel equations can be solved by means of motion planning methods using feedforward control. 
Different from \cite{Fischer2020} the solution of these kernel equations is a much more challenging problem. This is due to the fact that it cannot be reformulated as a set-point change, which was the case in \cite{Fischer2020}. In particular, a transition problem with a non-equilibrium initial state and a vanishing end state has to be solved, resulting in an involved motion planning problem. For this, only a few results are available in the literature for parabolic and hyperbolic PDEs (see, e.g., \cite{Laroche2000,Woittennek2010a}), which, however, cannot be used to solve the fault diagnosis kernel equations. Hence, a new systematic solution procedure is proposed by combining backstepping with motion planning methods based on differential parametrizations of the system variables in terms of a parametrizing variable. So far, this idea was only formulated for scalar parabolic PDEs with varying coefficients in \cite{Meurer2009a}. Hence, the paper also contains a new and systematic result for the motion planning of hyperbolic ODE-PDE systems. To be more precise, the PDE subsystem is mapped into backstepping coordinates by making use of the results in \cite{Hu2019}. On the basis of the simple structure of the resulting target system, a parametrization of the system variables for the ODE-PDE system in terms of a parametrizing variable is determined. With this, the trajectory planning problem for the kernel equations can be traced back to a significantly simpler trajectory planning for the parametrizing variable. For the latter problem, a new constructive design procedure is presented to take the non-equilibrium initial state into account. As a result, an easy verifiable condition for the solvability of the fault diagnosis kernel equations and thus for fault detectability is found.

\subsection{Organization}
The following section presents the considered fault diagnosis problem. Afterwards, the fault diagnosis equation is derived in Section \ref{sec:faultdiagnosisEquation}. The systematic solution of the kernel equations is shown in Section \ref{sec:kernelSolution}. Finally, the method is demonstrated for a $4\times 4$ heterodirectional hyperbolic system coupled with a second order ODE subject to constant, ramplike and sinusoidal faults as well as sinusoidal and bounded disturbances.

\section{Problem formulation} \label{sec:problem}
Consider the \emph{faulty general linear heterodirectional hyperbolic ODE-PDE system}
\begin{subequations}\begin{align}
	x'(z,t) &= \Gamma(z) \dot{x}(z,t) + A(z)x(z,t)
	+ \rlap{$A_0(z) x^{\n}(0,t)$}
\notag\\&\quad 
+ \INT*{0}{z}{D(z, \zeta) x(\zeta, t)}{\zeta} + H_1(z) w(t)
\notag\\&\quad 
+ B_1(z) u(t) + E_1(z) f(t) + G_1(z) d(t)
\label{eq:sys:pde}\\
	x^{\p}(0,t) &= Q_0 x^{\n}(0,t) + H_2 w(t) + B_2 u(t)
	\notag\\&\quad  
	+ E_2f(t) + G_2 d(t)
\label{eq:sys:bc0}\\
	x^{\n}(1,t) &= Q_1 x^{\p}(1,t) + B_3 u(t) + E_3 f(t)
	\notag\\&\quad
		 + G_3 d(t)
\label{eq:sys:bc1}
	\\
	\dot{w}(t) &= F w(t) + L_2 x^{\n}(0,t) + B_4 u(t)\rlap{$ + E_4 f(t)$}
	\notag\\&\quad 
	 + G_4 d(t)
\label{eq:sys:ode}\\
	y(t) &= x^{\n}(0,t) + E_5 f(t) + G_5 d(t)
	\label{eq:sys:y}
\end{align}\label{eq:sys}\end{subequations}
with $n_x$ coupled \emph{transport PDEs} \eqref{eq:sys:pde} on $(z,t)\in(0,1) \times \R[+]$ and an ODE \eqref{eq:sys:ode} on $t\in\R[+]$. The system \eqref{eq:sys} has the distributed state $x(z,t)\in\R[n_x]$, the lumped state $w(t)\in\R[n_w]$, the known input $u(t)\in\R[n_u]$, the known output $y(t)\in\R[n_\n]$, the unknown fault $f(t)\in\R[n_f]$ and the unknown disturbance $d(t)\in\R[n_d]$. The transport behavior of the distributed state in \eqref{eq:sys:pde} is specified by $\Gamma(z) = \diag*{\gamma_1(z), \gamma_2(z), \dots, \gamma_{n_x}(z)}\in\R[n_x \times n_x]$ with $\gamma_i\in C^1[0,1]$, $i=1,\dots,n_x$. Without loss of generality, it is assumed that the nonvanishing transport velocities $\lambda_i(z) = \frac{1}{\gamma_i(z)}$, $i=1,\dots,n_x$, are sorted descending, i.e., 
\begin{subequations}\begin{align}
&\lambda_1(z) > \lambda_2(z) > \dots > \lambda_{n_\n}(z) > 0
\\
&0 > \lambda_{n_\n+1}(z) >  \lambda_{n_\n+2}(z) > \dots > \lambda_{n_x}(z).
\end{align}\label{eq:lambdas}\end{subequations}
In view of \eqref{eq:lambdas}, the state $x(z,t)$ can be split into 
\begin{subequations}
\begin{align}
x^{\n}(z,t) &= J_\n x(z,t)\in\R[n_\n]
\\
x^{\p}(z,t) &= J_\p x(z,t)\in\R[n_\p]
\end{align}
with
\begin{align}
	J_\n &= \MATRIX{I_{n_\n} & 0}\in\R[n_\n \times n_x]
	\\
	J_\p &= \MATRIX{0 & I_{n_\p}}\in\R[n_\p\times n_x]
\end{align}\label{eq:states}\end{subequations}
where $n_\p + n_\n= n_x$. Note, that \eqref{eq:states} implies
\begin{align}
	x(z,t) &= J_\p^\transpose x^{\p}(z,t) + J_\n^\transpose x^\n(z,t).
	\label{eq:states:x}
\end{align}
According to \eqref{eq:lambdas} and \eqref{eq:states} the state $x^\n(z,t)$ describes the propagation in the negative direction of the spatial coordinate and $x^{\p}(z,t)$ the propagation in positive direction.
Consequently, the PDE subsystem \eqref{eq:sys:pde}--\eqref{eq:sys:bc1} is a \emph{heterodirectional hyperbolic system}.
Furthermore, the entries of $A(z)=[A_{ij}(z)]\in\R[n_x\times n_x]$ satisfy $A_{ij}\in C^1[0,1]$, $i,j=1,\dots,n_x$ and $A_{ii}(z) = 0$, $z\in[0,1]$, $i=1,\dots,n_x$. The latter means no loss of generality, as this form can be achieved by a change of coordinates (see, e.g., \cite{Hu2016}).
The remaining matrices for the boundary coupling and the integral term in \eqref{eq:sys:pde} are $A_0\in(C^1[0,1])^{n_x\times n_\p}$ as well as $D(z,\zeta)\in(C^1([0,1]^2))^{n_x \times n_x}$.
With \eqref{eq:sys:bc0} and \eqref{eq:sys:bc1} defined on $t\in\R[+]$, the boundary conditions (BCs) are specified by $Q_0\in\R[n_\p \times n_\n]$ and $Q_1\in\R[n_\n\times n_\p]$. 
The dynamics of the ODE \eqref{eq:sys:ode} are characterized by $F\in\R[n_w\times n_w]$ and its coupling with the PDE via $L_2\in\R[n_w\times n_\n]$, $H_1\in(L_2(0,1))^{n_x\times n_w}$ as well as $H_2\in\R[n_\p\times n_w]$. 
The influence of the input $u(t)$, the fault $f(t)$ and the disturbance $d(t)$ on \eqref{eq:sys} are defined by $B_1\in(L_2(0,1))^{n_x\times n_u}$, $E_1\in(L_2(0,1))^{n_x\times n_f}$, $G_1\in(L_2(0,1))^{n_x\times n_d}$, real valued matrices $B_i$, $i=2,3,4$, $E_j$ and $G_j$, $j=2,3,4,5$ of appropriate dimensions.
The initial conditions (ICs) $x(z,0)\in\R[n_x]$, $z\in[0,1]$ and $w(0)\in\R[n_w]$ of \eqref{eq:sys} are unknown.

Different from the usual representation (see, e.g., \cite{Hu2016}), the PDE \eqref{eq:sys:pde} is solved for $x'(z,t)$, which is always possible. As shown in the following, this special form is well suited to solve the considered fault diagnosis problems.

With the setup \eqref{eq:sys} any \emph{additive fault} can be taken into account. This includes actuator, sensor as well as component faults. To be particular, a fault $f_i(t)$ as $i$th component of $f(t)$ that affects the $j$th input $u_j(t)$ with the corresponding input matrix $B_k$ is modelled by $E_k = B_k e_{j,n_u}\,e_{i,n_f}^\transpose$, with $e_{\iota,\nu}$, as $\iota$th unit vector in $\R[\nu]$ and $k=1,\dots,4$. Accordingly, a sensor fault $f_i(t)$ in the $j$th output is modeled by $E_5 = e_{j,n_y}\,e_{i,n_f}^\transpose$. If the entries in $E_i$, $i=1,\dots,5$, do not correspond to an input or output, then the corresponding fault is called a \emph{component fault} (see, e.g., \cite{Chen2012}). 

The unknown disturbance $d(t)$ is composed of a deterministic component $\tilde{d}(t)\in\R[n_{\tilde{d}}]$ and bounded component $\bar{d}(t)\in\R[n_{\bar{d}}]$, i.e.,
\begin{align}
d(t) &= \tilde{G} \tilde{d}(t) + \bar{G} \bar{d}(t)
\label{eq:disturbance}
\end{align}
with $\tilde{G}\in\R[n_d\times n_{\tilde{d}}]$ and $\bar{G}\in\R[n_d\times n_{\bar{d}}]$. The bounded part $\bar{d}(t)=[\bar{d}_i(t)]$ satisfies
\begin{align}
\abs{\bar{d}_i(t)} \leq \delta_i, \quad i=1,\dots,n_{\bar{d}}
\label{eq:disturbance:bounds}
\end{align}
with the known upper bound $\delta=[\delta_i]\in\R[n_{\bar{d}}]$ and the deterministic part $\tilde{d}(t)$ is the solution of a \emph{signal model}
\begin{subequations}\begin{align}
	\dot{v}_d(t) &= S_d v_d(t), && t > 0
	\label{eq:signalModel:dist:ode}
	\\
	\tilde{d}(t) &= \tilde{R}_d v_d(t), && t \geq 0
	\label{eq:signalModel:dist:d}
\end{align}\label{eq:signalModel:dist}\end{subequations}
with $v_d(t)\in\R[n_{v_d}]$, the known matrices $S_d\in\R[n_{v_d}\times n_{v_d}]$, $\tilde{R}_d\in\R[n_{\tilde{d}}\times n_{v_d}]$ and unknown IC $v_d(0)\in\R[n_{v_d}]$.

Similar to, $\tilde{d}(t)$, the fault $f(t)$ is also a solution of a signal model
\begin{subequations}\begin{align}
	\dot{v}_f(t) &= S_f v_f(t), && 
	t > 0
	\label{eq:signalModel:fault:ode}
	\\
	f(t) &= \tilde{R}_f v_f(t), && t \geq 0
	\label{eq:signalModel:fault:f}
\end{align}\label{eq:signalModel:fault}\end{subequations}
with $v_f(t)\in\R[n_{v_f}]$, known $S_f\in\R[n_{v_f}\times n_{v_f}]$ and known $\tilde{R}_f\in\R[n_f \times n_{n_f}]$. In order to model the occurrence of faults, the IC $v_f(t_i)$, $i\in\mathbb{N}_0$, may change on unknown time instants $t_i$, which gives rise to piecewise defined solutions of \eqref{eq:signalModel:fault}. Furthermore, it is assumed that the time intervals, in which the fault changes is uniformly lower bounded, i.e., $t_{i+1} - t_i > \Delta t \in\R^+$ must hold.

For the following computations, it is suitable to combine both signal models \eqref{eq:signalModel:dist} and \eqref{eq:signalModel:fault} into the extended signal model
\begin{subequations}\begin{align}
	\dot{v}(t) &= S v(t), && 
	t > 0
	\label{eq:signalModel:ode}
	\\
	f(t) &= R_f v(t), && t \geq 0
	\label{eq:sys:f}
	\\
	\tilde{d}(t) &= R_d v(t), && t \geq 0
	\label{eq:sys:d}
\end{align}\label{eq:signalModel}\end{subequations}
where $v = \col{v_f, v_d}\in\R[n_v]$, $n_v = n_{v_f} + n_{v_d}$, $S = \diag*{S_f,\,S_d}$, $R_f = [\tilde{R}_f\enspace 0]$ and $R_d = [0\enspace \tilde{R}_d]$. In order to model steplike, polynomial and trigonometric faults $f(t)$ and disturbances $\tilde{d}(t)$, the spectra of the matrices $S_f$ and $S_d$ satisfy $\sigma(S_f)$, $\sigma(S_d)\subset j\R$ and are not required to be diagonalizable. 
The signal model \eqref{eq:signalModel:ode} describes only the specific type of signals, but their actual form (e.g., amplitudes and phases in case of trigonometric signals) is defined by the unknown IC $v(t_i)\in\R[n_v]$. Hence, a large class of commonly occurring faults and disturbances can be considered.

In the paper, the following fault diagnosis problems are solved:
\begin{enumerate}
	\item \emph{fault detection}: detection of the occurrence of a fault $f(t)$,
	\item \emph{fault isolation}: independent detection of each fault $f_i(t)$, $i=1,\dots, n_f$ and
	\item \emph{fault identification}: determination of $f(t)$.
\end{enumerate}
Note that fault identification is only possible for $\bar{d}(t)\equiv 0$.
For the case $\bar{d}(t)\not\equiv 0$ a \emph{fault estimation} is investigated, which yields bounds for the estimation error of fault $f(t)$.

\section{Fault diagnosis equation}\label{sec:faultdiagnosisEquation}
The fault diagnosis equation is based on an explicit expression for the fault. To determine it, an input-output equation is derived, which is based on the application of integral transformations to the faulty system.

\subsection{Derivation of the input-output equation}
Consider the integral transformation for the PDE \eqref{eq:sys:pde}
\begin{subequations}\begin{align}\label{eq:transformation:M}
	\mathcal{M}[h](t) &= \INT{0}{1}{\INT{0}{T}{M^\transpose(z,\tau) h(z,t-\tau)}{\tau}}{z}
	\notag\\
	&= \innerp{M & h(t)}_{\Omega,\Time},
	\quad h(z,t)\in\R[n_x],
\end{align}
for the ODE \eqref{eq:sys:ode}
\begin{align}
	\mathcal{P}[h](t) &= \INT{0}{T}{P^\transpose(\tau) h(t-\tau)}{\tau}
	\notag\\
	&= \innerp{P & h(t)}_{\Time},
	\quad h(t)\in\R[n_w]
	\label{eq:transformation:P}
\end{align}
and for the signal model \eqref{eq:signalModel}
\begin{align}
	\mathcal{Q}[h](t) &= \innerp{Q & h(t)}_{\Time},
	\quad
	 h(t)\in\R[n_v].
	\label{eq:transformation:Q}
\end{align}\label{eq:transformation}\end{subequations}
Note that due to the delay character of the hyperbolic system \eqref{eq:sys}, the detection time $T$ has to satisfy $0 < T_{0} < T < \Delta t$. The lower bound $T_0$ is a system property, which will be specified later and the upper bound is introduced in Section \ref{sec:problem}. 
Since the time integrations for $h(t)$ are given by the sliding interval $\Time_{t} = [t-T, t]$, the transformations \eqref{eq:transformation} can be implemented online.
The kernels $M(z,\tau)\in\R[n_x\times n_f]$, $P(\tau)\in\R[n_w\times n_f]$ and $Q(\tau)\in\R[n_v\times n_f]$ on $z\in\Omega=[0,1]$, $\tau\in\Time=[0,T]$ have to be determined accordingly to the fault diagnosis problem. To this end, $M(z,\tau)$ and $P(\tau)$ are determined to obtain the \emph{input-output equation}
\begin{align}
\innerp{M_E & f(t)}_{\Time} 
&=
- \innerp{ N & y(t)}_{\Time}
- \innerp{M_B & u(t)}_{\Time}
\notag\\&\qquad
- \innerp{ M_{\tilde{G}} & \tilde{d}(t)}_{\Time}
- \innerp{ M_{\bar{G}} & \bar{d}(t)}_{\Time}
\label{eq:io}\end{align}
depending only on $f(t)$, $y(t)$, $u(t)$ and $d(t)$ after applying \eqref{eq:transformation:M} and \eqref{eq:transformation:P} to \eqref{eq:sys:pde} respectively \eqref{eq:sys:ode}. The matrices appearing in \eqref{eq:io} are
\begin{subequations}\begin{align}
	M_E(\tau) &= \innerp{E_1 & M(\tau)}_{\Omega} + E_2^\transpose M_\p(0,\tau) 
	\notag\\&\quad
	- E_3^\transpose M_\n(1,\tau) + E_4^\transpose P(\tau) - E_5^\transpose N(\tau)
	\label{eq:ME}
	\\
	M_B(\tau) &= \innerp{B_1 & M}_\Omega + B_2^\transpose M_\p(0) - B_3^\transpose M_\n(1)
	\notag\\&\quad
	+ B_4^\transpose P
	\label{eq:MB}
	\\
	M_{\tilde{G}}(\tau) &= \tilde{G}^\transpose M_G(\tau)
	\label{eq:MG:tilde}
	\\
	M_{\bar{G}}(\tau) &= \bar{G}^\transpose M_G(\tau),
\end{align}
in which $N(\tau)\in\R[n_\n]$, is a degree of freedom that will be used to determine the kernel $M(z,\tau)$. Furthermore, 
\begin{align}
	M_G(\tau) &= \innerp{G_1 & M(\tau)}_{\Omega} + G_2^\transpose M_\p(0,\tau) 
	\notag\\&\quad- G_3^\transpose M_\n(1,\tau) + G_4^\transpose P(\tau) - G_5^\transpose N(\tau),
	\label{eq:MG}
\end{align}\end{subequations}
$M_\p(z,\tau) = J_\p M(z,\tau)$, $M_\n(z,\tau) = J_\n M(z,\tau)$ are introduced and $\innerp{\cdot & M(\tau)}_\Omega$ denotes an integration \wrt $z$ on $\Omega$.
In Appendix \ref{sec:ioEquation}, it is shown that for this, the kernels $M(z,\tau)$ and $P(\tau)$ must satisfy the \emph{kernel equations}
\begin{subequations}
\begin{flalign}
	M'(z,\tau) &= - \Gamma(z) \dot{M}(z,\tau) - A^\transpose(z) M(z,\tau) 
	\notag\\&\qquad - \mathcal{D}^*[M(\tau)](z)
	\label{eq:requirement:pde}
	\\
	M_\n(0,\tau)
	&=
	- Q_0^\transpose M_\p(0,\tau)
	- \innerp{ A_0 & M(\tau) }_\Omega 
	\notag\\&\qquad
	- L_2^\transpose P(\tau)
	+ N(\tau)
	\label{eq:N}
	\\
	M_\p(1,\tau) &= - Q_1^\transpose M_\n(1,\tau)
	\label{eq:requirement:BC:1}
	\\
	\dot{P}(\tau) 
	&=F^\transpose P(\tau) +
	H_2^\transpose M_\p(0,\tau)
	\notag\\&\quad + \innerp{H_1 & M(\tau)}_\Omega
	\label{eq:requirement:ode:P}
	\\
	\left. M(z,\tau) \right|_{\tau\in\{0,T\}} &= 0,\quad \forall z\in\Omega
	\label{eq:requirement:M:IE}
	\\
	\left.P(\tau)\right|_{\tau\in\{0,T\}} &= 0, &
	\label{eq:requirement:P:IE}
\end{flalign}\label{eq:kernel:faultdiagnosis:MP}\end{subequations}
where \eqref{eq:requirement:pde} is defined on $(z,\tau)\in(0,1)\times(0,T)$, \eqref{eq:N}--\eqref{eq:requirement:ode:P} on $\tau\in(0,T)$, $\dot{M}(z,\tau) = \dtau M(z,\tau)$ and 
\begin{align}
	\mathcal{D}^*[M(\tau)](z) =  \INT{z}{1}{ D^\transpose(\zeta, z) M(\zeta,\tau) }{\zeta}.
	\label{eq:Dstar}
\end{align}

\subsection{Fault identification}
In the following, the fault identification problem is solved for $\bar{d}(t)\equiv 0$. Inserting \eqref{eq:sys:f} and \eqref{eq:sys:d} in \eqref{eq:io} as well as assuming $\bar{d}(t)\equiv 0$ yields
\begin{align}
	&\innerp{R_f^\transpose M_E + R_d^\transpose M_{\tilde{G}} & v(t)}_{\Time} \notag\\
	&\quad = - \innerp{ N & y(t)}_{\Time}
	- \innerp{M_B & u(t)}_{\Time}.
	\label{eq:io:tmp}\end{align}
In order to solve \eqref{eq:io} for the fault $f(t)$, apply the transformation \eqref{eq:transformation:Q} to \eqref{eq:signalModel:ode}, substitute $\dt Q(t-\tau) = -\dtau Q(t-\tau)$ and utilize an integration by parts. This leads to
\begin{subequations}\begin{flalign}
	\innerp{\dot{Q} - S^\transpose Q& v(t)}_{\Time} &= Q^\transpose(T) v(t-T) - Q^\transpose(0) v(t)\hspace{-3em} &
	\label{eq:derivation:Q:d}
\end{flalign}
giving with
\begin{align}
	Q^\transpose(T) &= 0 &&\text{and}& Q^\transpose(0) &= R_f
	\label{eq:requirement:Q:IE}
\end{align}\label{eq:requirement:Q}\end{subequations}
the result
\begin{align}
	f(t) &= R_f v(t) = - \innerp{\dot{Q} - S^\transpose Q& v(t)}_{\Time}
	\label{eq:derivation:f}
\end{align}
(see \eqref{eq:sys:f}). Hence, imposing 
\begin{align}
\dot{Q}(\tau) - S^\transpose Q(\tau) = R_f^\transpose M_E(\tau) + R_d^\transpose M_{\tilde{G}}(\tau)
\label{eq:requirement:Q:ode}
\end{align}
and inserting this in \eqref{eq:io:tmp} yields
\begin{align}
	f(t) = \innerp{ N & y(t)}_{\Time} + \innerp{M_B & u(t)}_{\Time},\quad t\in\Time_{t_i}.
	\label{eq:io:identification}
\end{align}	
Since \eqref{eq:io:identification} only depends on known signals $y(t)$ and $u(t)$, the fault $f(t)$ is directly determined so that both fault isolation and identification are achieved.
This result is summarized in the next theorem.
\begin{thm}\label{thm:identification}\textbf{\emph{(Fault identification)}}
	Assume $\bar{d}(t) \equiv 0$, and that $f(t)$ as well as $\tilde{d}(t)$ are described by \eqref{eq:signalModel}. If $M(z,\tau)$, $P(\tau)$ and $Q(\tau)$ satisfy the kernel equations \eqref{eq:kernel:faultdiagnosis:MP}, \eqref{eq:requirement:Q:IE}, \eqref{eq:requirement:Q:ode}, then the fault $f(t)$ occurring at $t_i$, $i\in\mathbb{N}_0$ is identified by \eqref{eq:io:identification} for $t\geq t_i+T$.
\end{thm}
Note that the fault is identified not before the end of the detection interval $[t_i, t_i+T]$, i.e., at $t = t_i+T$.

\subsection{Fault detection}
If the disturbance $d(t)$ is subject to a bounded disturbance $\bar{d}(t)$, then only fault detection is possible. This is achieved by introducing a \emph{threshold}. 

Taking $\bar{d}(t)$ in \eqref{eq:io:identification} into account yields
\begin{align}
	f(t) = \tilde{f}(t) + \bar{f}(t)
	\label{eq:io:detection}
\end{align}
with the known part
\begin{align}
	\tilde{f}(t) &= \innerp{ N & y(t)}_{\Time} + \innerp{M_B & u(t)}_{\Time}
	\label{eq:fTilde}
\end{align}
and the unknown part
\begin{align}
	\bar{f}(t) &= \innerp{M_{\bar{G}} & \bar{d}(t)}_{\Time}.
	\label{eq:fault:bounded}
\end{align}
The absolute value of the detection error caused by $\bar{d}(t)$ for each fault $f_i(t)$ is
\begin{align}
	\abs{f_i(t) - \tilde{f}_i(t)} = \abs{\bar{f}_i(t)}, \quad i = 1,\dots, n_f
	\label{eq:fault:absoluteError}
\end{align}
with 
\begin{align}
	\bar{f}_i(t) &= \innerp{m_{\bar{G},i} & \bar{d}(t)}_{\Time}
	\label{eq:fault:bounded:i}
\end{align}
and $m_{\bar{G},i}(\tau) = M_{\bar{G}}(\tau)e_{i,n_f}$ in view of \eqref{eq:fault:bounded}. Hence, an upper bound for guaranteed fault detection can be found by estimating $\abs{\bar{f}_i(t)}$. To this end, use the integral representation of $\innerp{\cdot & \cdot}_{\Time}$, to obtain for \eqref{eq:fault:absoluteError} the upper bound
\begin{align}
	\abs{\bar{f}_i(t)} \leq \INT{0}{T}{ \abs{m_{\bar{G},i}^\transpose(\tau) \bar{d}(t-\tau)}}{\tau}.
	\label{eq:derivation:identifiaction}
\end{align}
In order to be able to use \eqref{eq:disturbance:bounds} in \eqref{eq:derivation:identifiaction}, the vector of absolute values $\abs{h(t)} = \col*{\abs{h_1(t)}, \abs{h_2(t)}, \dots, \abs{h_\nu(t)}) \in(\R_0^+}^\nu$, $h(t)\in\R[\nu]$ is introduced. Thus, 
\begin{align}
	\abs{m_{\bar{G},i}^\transpose(\tau) \bar{d}(t-\tau)} 
	\leq 
	\abs{ m_{\bar{G},i}(\tau) }^\transpose \abs{\bar{d}(t-\tau)}
	\label{eq:derivation:estimation:1}
\end{align}
follows for the integrand in \eqref{eq:derivation:identifiaction}. Hence, \eqref{eq:derivation:estimation:1} is bounded by
\begin{align}
	\abs{ m_{\bar{G},i}^\transpose(\tau) \bar{d}(t-\tau)} 
	\leq 
	\abs{ m_{\bar{G},i}(\tau) }^\transpose \delta
	\label{eq:derivation:estimation:2}
\end{align}
in view of \eqref{eq:disturbance:bounds}. Consequently, the threshold
\begin{align}
	f_{B,i} &= \INT{0}{T}{\abs{ m_{\bar{G},i}(\tau) }^\transpose }{\tau} \delta,
	\quad i=1,\dots,n_f
	\label{eq:threshold}
\end{align}
follows from \eqref{eq:derivation:identifiaction} and \eqref{eq:derivation:estimation:2}, which can be computed.
The next theorem summarizes the obtained results.
\begin{thm}\label{thm:detection}\textbf{\emph{(Fault detection)}}
	Let $\bar{d}(t)$ satisfy \eqref{eq:disturbance:bounds}. Then, a fault $f_i(t)$, $i=1,\dots,n_f$, occurring at $t_j$, $j\in\mathbb{N}_0$ is detected at $t\geq t_j+T$ if the threshold $f_{B,i}$ in \eqref{eq:threshold} is exceeded by the $i$th component $\tilde{f}_{i}(t)$ of $\tilde{f}(t)$ in \eqref{eq:fTilde}, i.e.,
	\begin{align}
	\abs{\tilde{f}_i(t)} > f_{B,i}, \qquad i=1,\dots,n_f.
	\label{eq:detection}
	\end{align}
	An estimate for the fault is given by
	\begin{align}
	\tilde{f}_i(t) - f_{B,i} \leq f_i(t) \leq \tilde{f}_i(t) + f_{B,i}. 
	\label{eq:estimation}
	\end{align}
\end{thm}
The proof of this theorem follows from \eqref{eq:fault:absoluteError} and the estimates \eqref{eq:derivation:identifiaction} and \eqref{eq:threshold}.

\section{Solution of the fault diagnosis kernel equations}\label{sec:kernelSolution}
In a first step, the kernel equations \eqref{eq:kernel:faultdiagnosis:MP}, \eqref{eq:requirement:Q:IE} and \eqref{eq:requirement:Q:ode} are rewritten in a suitable form. In particular, consider the matrices columnwise, i.e., $m_i = M e_{i,n_f}$, $n_i = N e_{i,n_f}$, $p_i = P e_{i,n_f}$, and $q_i = Q e_{i,n_f}$, $i=1,\dots,n_f$, where the arguments are omitted for convenience and use the spatial reversal $\bar{z}= 1-z$. Therein, the resulting quantities in $\bar{z}$ are denoted by adding an overline, e.g., $\bar{\Gamma}(\bar{z}) = \Gamma(1-\bar{z})$ with the diagonal elements $\bar{\gamma}_j(\bar{z}) = \gamma_j(1 - \bar{z})$, $j=1,\dots,n_x$. Furthermore, note that $\bar{m}_i'(\bar{z},\tau) = \partial_{\bar{z}}\bar{m}_i(\bar{z},\tau)$ and define $\mathcal{D}^*[m_i(\tau)](z) \rightarrow \INT{0}{\bar{z}}{\bar{D}^\transpose(\zeta, \bar{z}) \bar{m}_i(\zeta,\tau)}{\zeta}$. With this, one obtains
\begin{subequations}\begin{align}
	\bar{m}_i'(\bar{z},\tau) &= \bar\Gamma(\bar{z}) \dot{\bar{m}}_i(\bar{z},\tau) + \bar{A}^\transpose(\bar{z}) \bar{m}_i(\bar{z},\tau)
	\notag\\&\quad+ \INT*{0}{\bar{z}}{\bar{D}^\transpose(\zeta,\bar{z}) \bar{m}_i(\zeta,\tau) }{\zeta}
	\label{eq:faultDiagnosisKernel:pde}
	\\
	\bar{m}^{\p}_i(0,\tau) &= -Q_1^\transpose \bar{m}^{\n}_i(0,\tau)
	\\
	\bar{m}^{\n}_i(1,\tau) &= -Q_0^\transpose \bar{m}^{\p}_i(1,\tau) - \INT*{0}{1}{ \bar{A}_0^\transpose(\zeta) \bar{m}_{i}(\zeta,\tau)}{\zeta}
	\notag\\&\quad
		-L_2^\transpose J \eta_i(\tau) + n_i(\tau)
	\label{eq:faultDiagnosisKernel:bc1}
	\\
	\label{eq:faultDiagnosisKernel:ode}
	\dot{\eta}_i(\tau) &= \bar{F} \eta_i(\tau) + \INT*{0}{1}{\bar{B}_1(\zeta) \bar{m}_{i}(\zeta,\tau)}{\zeta} 
	\notag\\&\quad
	+ \bar{B}_2 \bar{m}^{\n}_i(0,\tau) + \bar{B}_3  \bar{m}^{\p}_i(1,\tau) 
	\notag\\&\quad 
	+ \bar{B}_4 n_i(\tau)
\end{align}\label{eq:faultDiagnosisKernel}\end{subequations}
where $\eta_i = \col{p_i & q_i}\in\R[n_\eta]$, $n_\eta=n_w + n_v$, and $J = \vect{I_{n_w} & 0}$ so that $p_i = J \eta_i$. Since $n_i(\tau)$ can be chosen freely, it is formally regarded as a boundary input. The appearing matrices in \eqref{eq:faultDiagnosisKernel:ode} result from the consideration of \eqref{eq:requirement:ode:P} and \eqref{eq:requirement:Q:ode} in view of \eqref{eq:ME}, \eqref{eq:MG:tilde} and read as
\begin{subequations}\begin{align}
	\bar{F} &= \MATRIX{ F^\transpose & 0 \\
	R_f^\transpose E_4^\transpose + R_d^\transpose \tilde{G}^\transpose G_4^\transpose & S^\transpose  }
	\\
	\bar{B}_1(\z) &= \MATRIX{\bar{H}_1^\transpose(\z) \\ R_f^\transpose \bar{E}_1^\transpose(\z) + R_d^\transpose \tilde{G}^\transpose \bar{G}_1^\transpose(\z)}
	\\
	\bar{B}_2 &= \MATRIX{0 \\ - R_f^\transpose E_3^\transpose  - R_d^\transpose \tilde{G}^\transpose G_3^\transpose }
	\\
	\bar{B}_3 &= \MATRIX{ H_2^\transpose \\ R_f^\transpose E_2^\transpose + R_d^\transpose \tilde{G}^\transpose G_2^\transpose}
	\\
	\bar{B}_4 &= \MATRIX{ 0 \\ - R_f^\transpose E_5^\transpose - R_d^\transpose \tilde{G}^\transpose G_5^\transpose}.
\end{align}\end{subequations}
The initial and end conditions following from \eqref{eq:requirement:M:IE}, \eqref{eq:requirement:P:IE} and \eqref{eq:requirement:Q:IE} are
\begin{subequations}\begin{align}
	\bar{m}_i(\bar{z},\tau)|_{\tau\in\{0, T\}} &= 0,  \quad \bar{z}\in\Omega
	\label{eq:faultDiagnosisKernel:IE:pde}
	\\
	\eta_i (0) &= \eta_i^0
	\label{eq:faultDiagnosisKernel:initial:ode}
	\\
	\eta_i(T) &= 0
	\label{eq:faultDiagnosisKernel:end:ode}
\end{align}\label{eq:faultDiagnosisKernel:IE}\end{subequations}
with $\eta_i^0 = \col*{0 & R_f^\transpose e_{i,n_f}}$.

Obviously, the solution of the kernel equations \eqref{eq:faultDiagnosisKernel} amounts to determining $n_i(\tau)$ such that a transition of $\bar{m}_i(z,\tau)$ and $\eta_i(\tau)$ from the initial to the end point in \eqref{eq:faultDiagnosisKernel:IE} is realized in finite time. With this, the fault detection is traced back to a controllability problem for the kernel equations. Note, that in general \eqref{eq:faultDiagnosisKernel:IE} cannot be reformulated as a setpoint change as in \cite{Fischer2020}. Nevertheless, a solution of \eqref{eq:faultDiagnosisKernel:IE} can still be constructed by representing it with a differential expression in terms of a parametrizing variable. The derivation of this expression is significantly facilitated by mapping the PDE subsystem \eqref{eq:faultDiagnosisKernel:pde}--\eqref{eq:faultDiagnosisKernel:bc1} into a target system of simpler structure. To this end, the substitution $\bar{z}\rightarrow z$ is used for notational convenience.

\subsection{Backstepping transformation}
In order to map \eqref{eq:faultDiagnosisKernel} into the \emph{target system}
\begin{subequations}\begin{align}
	\tilde{m}_i'(z, \tau) &= \bar{\Gamma}(z) \dot{\tilde{m}}_i(z,\tau) + \rlap{$\tilde{A}_0(z) \tilde{m}^{\n}_i(0,\tau)$}
	\label{eq:targetSystem:pde}
	\\
	\tilde{m}^{\p}_i(0,\tau) &= -Q_1^\transpose \tilde{m}^{\n}_i(0,\tau),
	&\qquad \tau\in(0,T)
	\label{eq:targetSystem:bc0}
	\\
	\tilde{m}^{\n}_i(1,\tau) &= \tilde{n}_i(\tau),
	&\tau\in(0,T)
	\label{eq:targetSystem:bc1}
	\\
	\dot{\eta}_i(\tau) &= \tilde{F} \eta_i(\tau) + \INT*{0}{1}{\tilde{B}_{1}(\zeta) \tilde{m}_{i}(\zeta,\tau) }{\zeta} \hspace{-4cm}
	\notag\\&\quad
	+ \tilde{B}_{2} \tilde{m}_{i}(0,\tau) + \tilde{B}_{3}  \tilde{m}_{i}(1,\tau) \hspace{-4cm}
	\label{eq:targetSystem:ode}
\end{align}\label{eq:targetSystem}\end{subequations}
with \eqref{eq:targetSystem:pde} defined on $(z,\tau)\in(0,1)\times(0,T)$ and \eqref{eq:targetSystem:ode} on $\tau\in(0,T)$, the invertible \emph{backstepping transformation}
\begin{align}
\tilde{m}_i(z,\tau) &= \bar{m}_i(z,\tau) - \INT{0}{z}{ K(z , \zeta) \bar{m}_i(\zeta,\tau) }{\zeta}
\notag\\
&= 
\mathcal{T}[\bar{m}_i(\tau)](z)
\label{eq:transformation:backstepping}
\end{align}
is introduced. For this, the \emph{kernel} $K(z,\zeta)\in\R[n_x \times n_x]$ has to be the solution of
\begin{subequations}\begin{flalign}
	&\Lambda(z) \dz K(z,\zeta) + \dzeta(K(z,\zeta)\Lambda(\zeta))
	\notag
	\\
	&\quad=
	- K(z,\zeta) \Lambda(z) \bar{A}^\transpose(z)
	+ \Lambda(z)\bar{D}^\transpose(\zeta,z)
	\hspace{-1cm}
	\notag\\&\quad\quad\quad
	- \INT*{\zeta}{z}{K(z,\bar{\zeta}) \Lambda(\bar{\zeta}) \bar{D}^\transpose(\zeta, \bar{\zeta}) }{\bar{\zeta}}
	\label{eq:kernelEquation:backstepping:pde}
	\\
	& \rlap{$K(z,0) \Lambda(0) (J_\n^\transpose - J_\p^\transpose Q_1^\transpose)
	= 
	\tilde{A}_0(z)$}
	\label{eq:kernelEquation:backstepping:A0}
	\\
	& \Lambda(z) K(z,z) - K(z,z) \Lambda(z)
	 = \Lambda(z)\bar{A}^\transpose(z)\hspace{-1cm}&
\end{flalign}
where $\Lambda(z) = \bar{\Gamma}^{-1}(z)$,  \eqref{eq:kernelEquation:backstepping:pde} is defined on $0 < \zeta < z < 1$ and
\begin{align}
	\tilde{A}_0(z) &= \MATRIX{\tilde{A}_{1}(z) \\ \tilde{A}_{2}(z)}
	\label{eq:A0}
\end{align}
is composed of $\tilde{A}_{2}(z)\in\R[n_\p\times n_\n]$ as well as the strictly lower triangular matrix $\tilde{A}_{1}(z)\in\R[n_\n\times n_\n]$ given by
\begin{align}
	\tilde{A}_{1}(z)
	 =\MATRIX{0 & \dots & \dots & 0 \\
	\tilde{A}_{1,21}(z) & \ddots & \ddots & \vdots \\
	\vdots & \ddots & \ddots & \vdots \\
	\tilde{A}_{1,n_\n 1}(z) & \dots & \tilde{A}_{1,n_\n n_\n-1}(z) & 0
	}
	.
\end{align}\label{eq:kernelEquation:backstepping}\end{subequations}
For the derivation of \eqref{eq:kernelEquation:backstepping} see \cite{Hu2019}. Therein, \eqref{eq:kernelEquation:backstepping} was investigated without the integral term. Since, the appearance of the latter term does not change the result in \cite{Hu2019}, it follows that \eqref{eq:kernelEquation:backstepping} has a piecewise $C^1$-solution.

The new input in \eqref{eq:targetSystem:bc1} is
\begin{align}\label{eq:targetSystem:n}
\tilde{n}_i(\tau) &= n_i(\tau) - Q_0^\transpose \bar{m}^{\p}_i(1,\tau) -L_2^\transpose J \eta_i(\tau)
\notag\\&\quad
- \INT*{0}{1}{\left( \bar{A}_0^\transpose(\zeta) + J_\n K(1,\zeta) \right) \bar{m}_{i}(\zeta,\tau) }{\zeta}
\end{align}
in view of \eqref{eq:faultDiagnosisKernel:bc1} and \eqref{eq:transformation:backstepping}.
The matrices in \eqref{eq:targetSystem:ode} result from applying, the \emph{inverse backstepping transformation}
\begin{align}
	\bar{m}_i(z,\tau) &= \tilde{m}_i(z,\tau) + \INT{0}{z}{K_I(z,\zeta) \tilde{m}_i(\zeta,\tau) }{\zeta}
	\notag\\&	= \mathcal{T}^{-1}[\tilde{m}_i(\tau)](z)
	\label{eq:backstepping:inverse}\end{align}
with $K_I(z,\zeta)\in\R[n_x \times n_x]$ (see \cite{Hu2019} for the details). In order to obtain \eqref{eq:targetSystem:ode}, \eqref{eq:faultDiagnosisKernel:bc1} is solved for $n_i(\tau)$ and the result is inserted in \eqref{eq:faultDiagnosisKernel:ode}. After utilizing \eqref{eq:backstepping:inverse} and changing the order of integration, \eqref{eq:targetSystem:ode} follows with the matrices
\begin{subequations}\begin{align}
	\label{eq:F:tilde}
	\tilde{F} &= \bar{F} + \bar{B}_4L_2^\transpose J
	\\
	\label{eq:targetSystem:ode:B1}
	\tilde{B}_{1}(z) 
	&= \bar{B}_{1}(z) 
	+ \INT*{z}{1}{ \bar{B}_{1}(\zeta) K_I(\zeta,z) }{\zeta}
	+ \bar{B}_2 J_\p K_I(1,z)
	\notag
	\\&\quad\notag
	+ \bar{B}_4 \Big( 
	(J_\n + K_0^\transpose J_\p) K_I(1,z)
	+ \bar{A}_0^\transpose(z) \hspace{-4cm}
	\\&\quad
	+ \INT*{z}{1}{\bar{A}_0^\transpose(\zeta) K_I(\zeta,z)}{\zeta} \Big)
	\\
	\tilde{B}_{2} &= \bar{B}_3J_\n
	\\
	\tilde{B}_{3} &= \bar{B}_2J_\p + \bar{B}_4(J_\n + Q_0^\transpose J_\p).
\end{align}\end{subequations}

\begin{figure}
	\centering
	\begin{tikzpicture}
	\newcommand{\distance}{1}
	\renewcommand*{\arraystretch}{1.2}
	\newcommand{\multilinebox}[1]{\begin{tabular}{@{}c@{}}#1\end{tabular}}
	\node[draw] (m) {\multilinebox{$\tilde{m}_i$-PDE \\ subsystem}};
	\draw[<-] (m.west) --node[above]{$\tilde{n}_i(\tau)$} ++(-\distance,0);
	\node[draw, right= 1.3cm of m] (eta) {\multilinebox{$\eta_i$-ODE \\ subsystem}};
	\draw[->] (m) --node[above]{$\tilde{m}_i(\tau)$} (eta);
	\draw[->] (eta.east) --node[above]{$\eta_i(\tau)$} ++(\distance,0);
	\end{tikzpicture}
	\caption{Cascade structure of the target system \eqref{eq:targetSystem}.}
	\label{fig:cascade}
\end{figure}
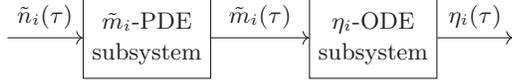

The target system \eqref{eq:targetSystem} is a ODE-PDE cascade with the boundary input $\tilde{n}_i(\tau)$, as depicted in Figure \ref{fig:cascade}. 
Additionally, the in-domain couplings $\bar{A}^\transpose(z)\bar{m}_i(z)$ and $\bar{D}^\transpose(\zeta,z) \bar{m}_i(\zeta)$ appearing in \eqref{eq:faultDiagnosisKernel:pde} are removed by the backstepping transformation \eqref{eq:transformation:backstepping}. Hence, the PDE subsystem \eqref{eq:targetSystem:pde}--\eqref{eq:targetSystem:bc1} is a cascade of transport equations. Consequently, the PDE and ODE subsystems can be considered successively for the derivation of the required differential expression.

\subsection{Parametrization for the solution of the PDE-subsystem}\label{sec:diffParam}
In order to construct an expression for $\tilde{m}_i(z,\tau)$, $\tilde{n}_i(\tau)$ and $\eta_i(\tau)$ in \eqref{eq:targetSystem}, a formal Laplace transform is applied. For this the correspondence $h(z,\tau)\laplace\check{h}(z,s)$ is utilized. In the sequel the argument $s$ is omitted for convenience, i.e., $\check{h}(z)=\check{h}(z,s)$. For a rigorous mathematical justification of the following see, e.g., \cite{Rudolph2008a,Woittennek2003}.

In a first step, an expression of $\tilde{m}_i(z,\tau)$ and $\tilde{n}_i(\tau)$ in terms of $\check{\tilde{m}}^{\n}_i(0)$ for the PDE subsystem \eqref{eq:targetSystem:pde}--\eqref{eq:targetSystem:bc1} is determined. To this end, apply the Laplace transform to \eqref{eq:targetSystem:pde} and \eqref{eq:targetSystem:bc0} to obtain the initial value problem
\begin{subequations}\begin{flalign}
	\check{\tilde{m}}_i'(z) &= s \bar{\Gamma}(z) \check{\tilde{m}}_i(z) + \tilde{A}_0(z) \check{\tilde{m}}^{\n}_i(0), \quad z\in(0,1)\hspace{-3em}&
	\label{eq:targetSystem:pde:s}
	\\
	\check{\tilde{m}}_i^\p (0) &= - Q_1^\transpose \check{\tilde{m}}_i^\n(0).
\end{flalign}\end{subequations}
The general solution of \eqref{eq:targetSystem:pde:s} is given by
\begin{align}
	\label{eq:solution:m}
	\check{\tilde{m}}_i(z) &= \check{\Phi}(z,0,s) V \check{\tilde{m}}^\n_i(0) 
	\notag\\&\quad 
	+ \INT*{0}{z}{\check{\Phi}(z,\zeta,s) \tilde{A}_0(\zeta)  }{\zeta} \check{\tilde{m}}^{\n}_i(0)
\end{align}
with $\check{\tilde{m}}_i(0) = V \check{\tilde{m}}^{\n}_i(0)$, $V = (J_\n^\transpose - J_\p^\transpose Q_1^\transpose)$ and the state transition matrix $\check{\Phi} : \Omega^2 \times \mathbb{C} \rightarrow \mathbb{C}^{n_x \times n_x}$. Due to the diagonal form of $\bar{\Gamma}(z)$ the latter can be explicitly determined as
\begin{subequations}\begin{align}
	&\check{\Phi}(z,\zeta,s) = \diag{\exp{s \theta_1(z, \zeta) }, \dots, \exp{s \theta_{n_x}(z, \zeta)}}
	\notag
\end{align}
where
\begin{align}
	\label{eq:theta}
	\theta_j(z,\zeta) &= \INT*{\zeta}{z}{ \bar{\gamma}_j(\bar{\zeta}) }{\bar{\zeta}}, \quad z,\zeta\in\Omega
\end{align}\end{subequations}
for $j=1,\dots,n_x$. Thus, in view of \eqref{eq:targetSystem:bc1} and \eqref{eq:solution:m} one can represent $\check{\tilde{m}}_i(z)$ and $\check{\tilde{n}}_i$ by 
\begin{subequations}\begin{align}
	\label{eq:solution:m:s}
	\check{\tilde{m}}_i(z) &= \check{\Psi}(z,s) \check{\tilde{m}}^{\n}_i(0), \quad z\in\Omega
	\\
	\check{\tilde{n}}_i &= J_\n \check{\Psi}(1,s) \check{\tilde{m}}_i^\n(0)
\end{align}\label{eq:parametrization:pde:s}\end{subequations}
with
\begin{align}
	\check{\Psi}(z,s) =
	\check{\Phi}(z,0,s) V 
	+ \INT*{0}{z}{\check{\Phi}(z,\zeta,s) \tilde{A}_0(\zeta) }{\zeta}.
	\label{eq:Psi:s}
\end{align}
In order to map \eqref{eq:solution:m} back into the time domain, \eqref{eq:Psi:s} is considered rowwise, i.e., $\check{\Psi}_j(z,s) = e_{j,n_x}^\transpose \check{\Psi}(z,s)$, $j=1,\dots,n_x$ in \eqref{eq:Psi:s}. Due to the diagonal form of $\check{\Phi}(z,\zeta,s)$ its $j$th row is $e_{j,n_x}^\transpose\check{\Phi}(z,\zeta,s) = e_{j,n_x}^\transpose \exp{s \theta_j(z, \zeta)}$ yielding
\begin{flalign}\label{eq:Psi:i:s}
	\check{\Psi}_j(z,s) &=
	e_{j,n_x}^\transpose V \exp{s\theta_j(z,0) }
	+ \INT*{0}{z}{ e_{j,n_x}^\transpose \tilde{A}_0(\zeta) \exp{s \theta_j(z, \zeta) } }{\zeta}\hspace{-1cm}&
\end{flalign}
for each row in \eqref{eq:Psi:s}. Utilizing the correspondence $\check{\Psi}_j(z,s)\check{h}\Laplace \Psi_j[h](z,\tau)$ with $\check{h}\Laplace h(\tau)\in\R[n_\n]$ for $\check{h}(s)\in\mathbb{C}^{n_\n}$, leads to the time-domain representation
\begin{align}
\label{eq:Psi:tau}
\Psi_j[h](z,\tau) &= e_{j,n_x}^\transpose V h(\tau+\theta_j(z,0))
\notag\\&\quad
+ \INT*{0}{z}{e_{j,n_x}^\transpose \tilde{A}_{0}(\zeta) h(\tau + \theta_j(z,\zeta) )}{\zeta}
\end{align}
of \eqref{eq:Psi:i:s} so that
\begin{align}
	\Psi[h](z,\tau) = \col{\Psi_1[h](z,\tau), \dots, \Psi_{n_x}[h](z,\tau)}.
	\label{eq:Psi:vect}
\end{align}
Using \eqref{eq:Psi:vect} in \eqref{eq:solution:m:s} the expressions
\begin{subequations}\begin{align}
\tilde{m}_{i}(z,\tau) &= \Psi\left[ \tilde{m}^{\n}_i(0) \right](z,\tau)
\label{eq:diffParam:pde:m}
\\
\tilde{n}_i(\tau) &= J_\n \Psi \left[ \tilde{m}^{\n}_i(0) \right](1,\tau)
\end{align}\label{eq:diffParam:pde}\end{subequations}
for $i=1,\dots,n_f$, in terms of $\tilde{m}_i(0,\tau)$ are obtained. Based on \eqref{eq:diffParam:pde} a solution for the cascade system \eqref{eq:targetSystem} can readily be constructed.

\subsection{Differential expression for the solution of the cascade system}
Applying the Laplace transform with $\d_\tau h(\tau) \laplace s \check{h}$ to \eqref{eq:targetSystem:ode} gives
\begin{align}
	\label{eq:eta:s:1}
	\big(s I_{n_\eta} - \tilde{F} \big) \check{\eta}_i &= 
	\INT*{0}{1}{\tilde{B}_{1}(\zeta) \check{\tilde{m}}_i(\zeta)}{\zeta}
	\notag\\&\quad
	+ \tilde{B}_{2} \check{\tilde{m}}_i(0)
	+ \tilde{B}_{3} \check{\tilde{m}}_i(1).
\end{align}
Inserting \eqref{eq:solution:m:s} yields
\begin{align}
	\label{eq:solution:eta:s}
	\big(s I_{n_\eta} - \tilde{F} \big) \check{\eta}_i &= 
	\big(
	\INT*{0}{1}{\tilde{B}_{1}(\zeta) \check{\Psi}(\zeta,s)}{\zeta}
	\notag\\&\quad
	+ \tilde{B}_{2} \check{\Psi}(0,s)
	+ \tilde{B}_{3} \check{\Psi}(1,s)
	\big)\check{\tilde{m}}^{\n}_i(0).
\end{align}
Obviously, $\check{\eta}_i$ and $\check{\tilde{m}}_i^\n(0)$ can be expressed by introducing the parametrizing variable $\check{\varphi}_i(s)\in\mathbb{C}^{n_\n}$ according to 
\begin{subequations}\begin{align}
	\label{eq:solution:eta:phi:s}
	\check{\eta}_i &= \adj\big(s I_{n_\eta} - \tilde{F} \big) 
	\big(
	\INT*{0}{1}{\tilde{B}_{1}(\zeta) \check{\Psi}(\zeta,s)}{\zeta}
	\notag\\&\quad
	+ \tilde{B}_{2} \check{\Psi}(0,s)
	+ \tilde{B}_{3} \check{\Psi}(1,s)
	\big)\check{\varphi}_i
	\\
	\label{eq:parametrization:m:n}
	\check{\tilde{m}}^{\n}_i(0) &= \det \big(sI - \tilde{F}\big) \check{\varphi}_i
\end{align}\label{eq:parametrization}\end{subequations}
in view of $(s I_{n_\eta} - \tilde{F}) \adj(s I_{n_\eta} - \tilde{F}) = \det(s I_{n_\eta} - \tilde{F})I_{n_\eta}$. 
This result can be mapped into the time domain by using 
\begin{align}
	\adj(sI_{n_\eta} - \tilde{F}) &= \sum_{i=0}^{n_\eta - 1} \tilde{F}_{i} s^i
	\label{eq:adj}
\end{align}
with $\tilde{F}_{i}\in\R[n_\eta \times n_\eta]$ to obtain the differential expression
\begin{flalign}
	\label{eq:diffParam:eta}
	\eta_i(\tau) &= \sum_{j=0}^{n_\eta-1} 
	\Big(
	\INT*{0}{1}{ \tilde{F}_{j} \tilde{B}_{1}(\zeta) \Psi\left[\d_\tau^j\varphi_i \right](\zeta,\tau) }{\zeta}
	\notag\\&\hspace{-.9cm}
	+ \tilde{F}_{j} \tilde{B}_{2} \Psi\left[\d_\tau^j\varphi_i \right](0,\tau)
	+ \tilde{F}_{j} \tilde{B}_{3} \Psi\left[\d_\tau^j\varphi_i \right](1,\tau)
	\Big)
\end{flalign}
for $\eta_i(\tau)$, $i=1,\dots,n_f$, in terms of $\d_\tau^j \varphi_i$, $j=0,1,\dots,n_\eta$. 
It remains to parametrize $\tilde{m}_i(z,\tau)$ and $\tilde{n}_i(\tau)$ in terms of $\varphi_i(\tau)$. For this, use 
\begin{align}
\det\big(sI - \tilde{F}\big) &= \sum_{i=0}^{n_\eta} \mu_i s^i
\label{eq:det:sIF}
\end{align}
with $\mu_i \in \R$, $\mu_{n_\eta}=1$ in \eqref{eq:parametrization:m:n} to obtain 
\begin{align}
	\label{eq:diffParam:m_n}
	\tilde{m}_i^\n(0,\tau) &= \sum_{j=0}^{n_\eta} \mu_j \d_\tau^j \varphi_i(\tau), \quad i=1,\dots,n_f&
\end{align}
in the time domain. Consequently, 
\begin{subequations}\begin{align}
	\tilde{m}_{i}(z,\tau) &= \sum_{j=0}^{n_\eta} \mu_j \Psi\left[ \d_\tau^j \varphi_i \right](z,\tau)
	\label{eq:diffParam:pde:m:phi}
	\\
	\tilde{n}_i(\tau) &= J_\n \sum_{j=0}^{n_\eta} \mu_j \Psi\left[ \d_\tau^j \varphi_i \right](1,\tau)
	\label{eq:diffParam:pde:n:phi}
\end{align}\label{eq:diffParam:pde:phi}\end{subequations}
follows from inserting \eqref{eq:diffParam:m_n} in \eqref{eq:diffParam:pde}.

\subsection{Reference trajectory planning}
To solve the initial-end value problem for \eqref{eq:targetSystem} the condition \eqref{eq:faultDiagnosisKernel:IE:pde} is mapped into 
\begin{align}
\tilde{m}_i(z,\tau)|_{\tau\in\{0,T\}} &= 0,\quad z\in\Omega
\label{eq:ie:backstepping}
\end{align}
via \eqref{eq:transformation:backstepping}. Then, the kernel $\tilde{m}_i(z,\tau)$ is obtained as a solution of the initial-end value problem \eqref{eq:targetSystem}, \eqref{eq:ie:backstepping}, \eqref{eq:faultDiagnosisKernel:initial:ode} and \eqref{eq:faultDiagnosisKernel:end:ode}. 
In the sequel this initial-end value problem is solved by the planning of a suitable reference trajectory $\varphi_{d,i}(\tau)$ for $\varphi_i(\tau)$. 

\subsubsection{Reference trajectory for the PDE subsystem}\label{sec:trj:pde}
In order to provide a systematic solution for the finite-time transition, the initial and end conditions for the PDE and ODE states are successively taken into account. At first, \eqref{eq:diffParam:pde:m} is used to formulate conditions on a reference trajectory $\tilde{m}_{d,i}^\n(0,\tau)$ for $\tilde{m}_{i}^\n(0,\tau)$ so that $\tilde{m}_{i}(z,\tau)$ satisfies \eqref{eq:ie:backstepping}. 
To this end, consider \eqref{eq:diffParam:pde:m} rowwise yielding
\begin{align}	\label{eq:mdi}
	&e_{j,n_x}^\transpose \tilde{m}_{d,i}(z,\tau) = 
	e_{j,n_x}^\transpose V \tilde{m}_{d,i}^\n(0, \tau+\theta_j(z,0))
	\notag\\&\quad
	+ \INT*{0}{z}{e_{j,n_x}^\transpose \tilde{A}_{0}(\zeta) \tilde{m}_{d,i}^\n(0, \tau + \theta_j(z,\zeta) )}{\zeta}
\end{align}
in view of \eqref{eq:Psi:tau}. Consequently, \eqref{eq:ie:backstepping} holds if $\tilde{m}_{d,i}^\n(0,\tau)$, $i=1,\dots,n_f$, satisfies
\begin{flalign}
	\tilde{m}^{\n}_{d,i}(0,\tau + \theta_j(z,\zeta))|_{\tau\in\{0,T\}} &= 0
	\label{eq:mn0:omega}
\end{flalign}
for $j=1,\dots,n_x$, on $0\leq \zeta \leq z \leq 1$. 
To ensure \eqref{eq:mn0:omega} by a suitable trajectory planning, the largest prediction time $\tau^+$ and delay time $\tau^-$ are required. They are given by the longest transportation time $\theta_{n_\n}(1,0)$ in the negative spatial direction and the longest transportation time $\abs{\theta_{n_\n+1}(1,0)}$ in the positive spatial direction (see \eqref{eq:lambdas} and \eqref{eq:theta}). Hence, 
\begin{align}
	\tau^\p &= \theta_{n_\n}(1,0) & \text{and} && \tau^- &= \abs{\theta_{n_\n+1}(1,0)}
	\label{eq:transportationTimes}
\end{align}
result.
Taking \eqref{eq:mn0:omega} and \eqref{eq:transportationTimes} into account, the reference trajectory for $\tilde{m}_{d,i}^\n(0,\tau)$ must be defined piecewise, i.e., 
\begin{flalign}
&\tilde{m}^\n_{d,i}(0,\tau) = 
\begin{cases}
0 &: \tau \in \Time_1 = [-\tau^-, \tau^+]
\\
\chi_i(\tau) &: \tau \in\Time_2 = (\tau^+, T - \tau^-)
\\
0 &: \tau\in\Time_3=[T - \tau^-,T+\tau^+],
\end{cases}\hspace{-3cm}&
\label{eq:mn0:piecewise}
\end{flalign}
where $\chi_i(\tau)\in\R[n_\n]$ and $\chi_i\not\equiv 0$ is a degree of freedom to be considered later. 
From Figure \ref{fig:mn0}, the condition $T - \tau^- > \tau^+$ is directly inferred, i.e., the detection time is lower bounded by 
\begin{align}
	T > \tau^+ + \tau^- = T_0.
\label{eq:Tmin}
\end{align}
\begin{figure*}
	\centering
%
%
		\DTLfetchsave{\tI}{dataTable}{key}{tau_mins}{value}
		\DTLfetchsave{\tII}{dataTable}{key}{tau_plus}{value}	
		\DTLfetchsave{\tIII}{dataTable}{key}{T_tau_min}{value}
		\DTLfetchsave{\tIV}{dataTable}{key}{T_tau_plus}{value}
		\DTLfetchsave{\tT}{dataTable}{key}{T}{value}	
		\newcommand{\gridUpper}{.8}
		\newcommand{\gridLower}{-.8}
		\centering
	\begin{tikzpicture}[trim axis left,trim axis right] 
		\begin{axis}[%
		name=plot,
		width=.95\textwidth,
		height= .8cm,
		scale only axis,
		xmajorgrids=false,
		ymajorgrids=false,
		ymin = -0.1,
		ymax = 0.7,
		xmin = -4,
		xmax = 34,
		axis lines=middle,
		inner axis line style={=>},
		clip = false,
		yticklabels={,,},
		ytickmax = 0.15,
		ytick=\empty,
		xlabel={$\tau $},
		xlabel style={anchor=south west},
		xtick = { 0.01, \tT},
		xticklabels = {$0$, $T$},
		xticklabel style = {anchor=north,fill=white,text opacity=1},
		ylabel style={anchor=east,outer sep=2pt,inner sep=0pt,fill=white,text opacity=1},
		ylabel={$\tilde{m}_{d,i}^-(0,\tau)$},
		]
		\addplot [color=blue,solid,very thick,restrict x to domain=\tII:\tIII,] table[x=tau,y=1]{data/mDesired.dat}node [pos=0.05, blue, above right] {$\chi_i(\tau)$};
		\addplot [color=red,solid,very thick,domain=\tI:\tII, samples=2, mark=*, mark size = 1pt, mark indices = {\numcoords}]{0};
		\addplot [color=red,solid,very thick,domain=\tIII:\tIV, samples=2, mark=*, mark size = 1pt, mark indices = {1}]{0};
		\newcommand{\gridLine}[1]{\draw[dashed, thick, gray] (axis cs: #1, \gridUpper) -- (axis cs:#1, \gridLower);}
		
		\gridLine{\tI}
		\gridLine{\tII}
		\gridLine{\tIII}
		\gridLine{\tIV}
		\newcommand{\xTickLabels}[3]{\node[#3,inner ysep=-.5ex, fill=white,opacity=1,text opacity=1]at (axis cs: #1,-.2){#2};}
		\xTickLabels{\tI}{$\tau^-$}{}
		\xTickLabels{\tII}{$\tau^+$}{}
		\xTickLabels{\tIII}{$T-\tau^-$}{}
		\xTickLabels{\tIV}{$T+\tau^+$}{}
		\newcommand{\arrowLower}{-.7}
		\newcommand{\raiseODE}{.3cm}
		\newcommand{\odePosition}{-1.1}
		
		\newcommand{\setCoordinate}[3]{\node(#3)[,outer sep=0pt,inner ysep=0pt,fill=white,text opacity=1] at (axis cs:#1,\arrowLower){#2};}
		
		\setCoordinate{\tI}{$\xi_i^0$}{nt1}
		\setCoordinate{\tII}{$\xi_i(\tau^+)$}{nt2}
		\setCoordinate{\tIII}{$\xi_i(T-\tau^-)$}{nt3}
		\setCoordinate{\tIV}{$\xi_i(T+\tau^+)$}{nt4}
		
		\draw[->,thick] (nt1) -- (nt2);
		\node at (axis cs:0,\odePosition) 
		{$\dot{\xi}_i(\tau) = A_\varphi \xi_i(\tau)$};
		\draw[->,thick] (nt2) -- (nt3);
		\node at (axis cs:15,\odePosition) 
		{$\dot{\xi}_i(\tau) = A_\varphi \xi_i(\tau) + B_\varphi \chi_i(\tau)$};
		\draw[->,thick] (nt3) -- (nt4);
		\node at (axis cs:\tT,\odePosition) 
		{$\dot{\xi}_i(\tau) = A_\varphi \xi_i(\tau)$};
		
		\newcommand{\setDomain}[3]{\node[fill=white,text opacity=1,xshift=#3,inner sep = 0pt,] at (axis cs:#1,.2){#2};}
				
		\setDomain{0}{$\Time_1$}{0cm}
		\setDomain{15}{$\Time_2$}{0cm}
		\setDomain{\tT}{$\Time_3$}{0cm}
		
		\newcommand{\setIE}[3]{\node(#3)[text opacity=1] at (axis cs:#1,1.2){#2};}
		
		\setIE{0}{$\red{\tilde{m}_i(z,0)=0,\eta_i(0) = \eta_i^0}$}{tm1}
		\setIE{\tT}{$\red{\tilde{m}_i(z,T)=0,\eta_i(T)=0}$}{tm2}
		\draw [->,blue] (tm1) -- (tm2);
		\end{axis}
	\end{tikzpicture}
	\caption{Visualization of \eqref{eq:mn0:piecewise} and the $\xi_i$-system.}
	\label{fig:mn0}
\end{figure*}%

\subsubsection{Reference trajectory for the cascade system}
A systematic approach to determine $\varphi_{d,i}(\tau)$ results from regarding \eqref{eq:diffParam:m_n} as ODE for $\varphi_{d,i}(\tau)$ with the input $\tilde{m}^\n_{d,i}(0,\tau)$. 
Hence, the ICs $\d_\tau^j \varphi_{d,i}(\tau)|_{\tau=-\tau^-}$, $j=0,1,\dots,n_\eta-1$ and the input $\tilde{m}^\n_{d,i}(0,\tau)$, i.e., $\chi_i(\tau)$ in \eqref{eq:mn0:piecewise}, have to be determined so that $\varphi_{d,i}(\tau)$ ensures the initial and end condition \eqref{eq:faultDiagnosisKernel:initial:ode}--\eqref{eq:faultDiagnosisKernel:end:ode} for $\eta_i(\tau)$ by evaluating \eqref{eq:diffParam:eta} at $\tau=0$ and $\tau=T$. The resulting solution has to satisfy $\varphi_{d,i}\in(C^{n_{\eta-1}}[- \tau^-, T+\tau^+])^{n_\n}$ with $\d_\tau^{n_\eta} \varphi_{d,i}$ existing in view of \eqref{eq:diffParam:m_n}.  This is facilitated by introducing the state
\begin{align}
	\xi_i(\tau) &= \MATRIX{\xi_{i,1}(\tau) \\ \xi_{i,2}(\tau) \\ \vdots \\ \xi_{i,n_\eta}(\tau)} \in \R[n_\eta n_\n],\quad i=1,\dots,n_f
	\label{eq:xi:state}
\end{align}
for \eqref{eq:diffParam:m_n} where $\xi_{i,j}(\tau) = \d_\tau^{j-1} \varphi_{d,i}(\tau) \in \R[n_\n]$, $j=1,\dots,n_\eta$, yielding
\begin{flalign}
\dot{\xi}_i(\tau) &= A_{\varphi} \xi_i(\tau) + B_\varphi \tilde{m}_{d,i}^\n(0,\tau),\enspace \tau\in(-\tau^-, T + \tau^+]
\hspace{-1em}
&
\label{eq:xi:ode}
\end{flalign}
for $i = 1,\dots, n_f$ with the matrices $A_{\varphi} = A_c \otimes I_{n_{\n}}$,
\begin{align}
A_c &= \MATRIX{
	0 & 1 & \dots & 0 \\
	\vdots & & \ddots \\
	0 & 0 & \dots & 1 \\
	-\mu_0 & - \mu_1 & \dots & -\mu_{n_\eta-1}
}
\label{eq:Ac}
\end{align}
and $B_\varphi = e_{n_{\eta},n_{\eta}} \otimes I_{n_\n}$. Therein, $A \otimes B$ is the \emph{Kronecker product}, i.e., $A\otimes B = [a_{ij}]\otimes B = [a_{ij}B]$ for matrices $A\in\R[m\times n]$ and $B\in\R[p\times q]$ of arbitrary dimension (see, e.g., \cite[Def. 7.1.2]{Bernstein2009}). Then, $\eta_i(\tau)$ can be represented in view of \eqref{eq:diffParam:eta} by
\begin{align}
\label{eq:diffParam:eta:xi}
\notag
\eta_i(\tau) &= \sum_{j=0}^{n_\eta-1} \Big(
\INT*{0}{1}{ \tilde{F}_{j} \tilde{B}_{1}(\zeta) 
	\Psi_\varphi[\xi_i](\zeta,\tau) }{\zeta}
\\&
+ \tilde{F}_{j} \tilde{B}_{2} \Psi_\varphi[\xi_i](0,\tau)
+ \tilde{F}_{j} \tilde{B}_{3} \Psi_\varphi[\xi_i](1,\tau)
\Big)
\end{align}
with $\Psi_\varphi[\xi_i](z,\tau) = \Psi [{U}_{j}\xi_i](z,\tau)$,
\begin{flalign}
	{U}_{j}\xi_i(\tau) &= \d_\tau^{j}\varphi_{d,i}(\tau)
	\label{eq:xi:output}
\end{flalign}
and ${U}_j = e_{j+1,n_\eta}^\transpose \otimes I_{n_\n}$ for $i=1,\dots,n_f$ as well as $j=0,\dots,n_\eta-1$.

According to \eqref{eq:mn0:piecewise} three cases have to be considered to determine $\tilde{m}^\n_{d,i}(0,\tau)$, which are illustrated in Figure \ref{fig:mn0}. In the first case $\tau\in\Time_1$ \eqref{eq:xi:ode} is autonomous. Thus, $\xi_i(\tau)$ is uniquely determined on $\Time_1$ by the IC $\xi_i(-\tau^-)=\xi_i^0$, which is a degree of freedom. It must be determined so that $\eta_i(0)$ resulting from \eqref{eq:diffParam:eta:xi} satisfies \eqref{eq:faultDiagnosisKernel:initial:ode}. To this end, use
\begin{align}
	\xi_i(\tau) &= \Phi_\varphi(\tau,-\tau^-)\xi_i^0,\quad \tau\in\Time_1
	\label{eq:xi:solution:1}
\end{align}
with $\Phi_\varphi(\tau,\tau_0) = \exp{A_{\varphi}(\tau-\tau_0)}$ in \eqref{eq:diffParam:eta:xi} and evaluate the result at $\tau=0$ giving
\begin{align}
	\label{eq:diffParam:eta:Phi:0}
	\eta_i(0) &= \sum_{j=0}^{n_\eta-1} \Big(
	\INT*{0}{1}{ \tilde{F}_{j} \tilde{B}_{1}(\zeta) 
		\Psi_\varphi\left[ \Phi_\varphi(-\tau^-)\xi_i^0 \right](\zeta,0) }{\zeta}
	\notag\\&\quad
	+ \tilde{F}_{j} \tilde{B}_{2} 
	\Psi_\varphi\left[ \Phi_\varphi(-\tau^-)\xi_i^0 \right](0,0)
	\notag\\&\quad
	+ \tilde{F}_{j} \tilde{B}_{3} 
	\Psi_\varphi\left[ \Phi_\varphi(-\tau^-)\xi_i^0 \right](1,0)
	\Big).
\end{align}
Utilizing $\Psi_\varphi[\Phi_\varphi(-\tau^-)\xi_i^0](\zeta,0) = \Psi_\varphi[\Phi_\varphi(-\tau^-)](\zeta,0) \xi_i^0$
in \eqref{eq:diffParam:eta:Phi:0}, the result
\begin{align}\label{eq:solution:eta:phi:xi:1}
	\eta_i^0 &= W_0 \xi_i^0
\end{align}
follows with
\begin{align}
	W_0 &= \sum_{j=0}^{n_\eta-1} 
	\Big(
	\INT*{0}{1}{ \tilde{F}_{j} \tilde{B}_{1}(\zeta) \Psi_\varphi\left[\Phi_\varphi(-\tau^-) \right](\zeta,0) }{\zeta}
	\notag\\&\quad
	+ \tilde{F}_{j} \tilde{B}_{2} 
	\Psi_\varphi\left[ \Phi_\varphi(-\tau^-) \right](0,0)
	\notag\\&\quad
	+ \tilde{F}_{j} \tilde{B}_{3} 
	\Psi_\varphi\left[ \Phi_\varphi(-\tau^-) \right](1,0)
	\Big).
\end{align}
If 
\begin{align}
	\rank W_0 &= \rank \MATRIX{W_0 & \eta_i^0},\quad i=1,\dots,n_f
	\label{eq:identifiability}
\end{align}
holds, then \eqref{eq:solution:eta:phi:xi:1} has a solution
\begin{align}
	\xi_i^0 &= W_0^\dagger \eta_i^0,
	\label{eq:xi:i0}
\end{align}
where $W_0^\dagger$ is the Moore-Penrose generalized inverse of $W_0$ (for details see, e.g., \cite[Prop. 6.1.7]{Bernstein2009}). 

In the second case $\xi_i(\tau)$ is determined on $\tau\in\Time_3$. Obviously, \eqref{eq:faultDiagnosisKernel:end:ode} is satisfied for the IC $\xi_i(T-\tau^-)=0$ implying
\begin{align}
	\xi_i(\tau) &= 0,\quad \tau\in\Time_3
	\label{eq:xi:solution:3}
\end{align}
in light of \eqref{eq:mn0:piecewise} and \eqref{eq:diffParam:eta:xi}.

In the last case it remains to determine $\xi_i(\tau)$ on $\Time_2$, which has to satisfy the initial and end condition
\begin{align}
	\xi_i(\tau^+) &= \Phi_\varphi(\tau^+, -\tau^-)\xi_i^0 && \text{and} & \xi_i(T-\tau^-) &= 0
	\label{eq:xi:transition}
\end{align}
implied by the previous cases. There exists an input $\chi_i(\tau)$ on $\tau\in\Time_2$ for \eqref{eq:xi:ode} (see \eqref{eq:mn0:piecewise}) such that \eqref{eq:xi:transition} holds provided \eqref{eq:xi:ode} is controllable. Then, this input takes the form 
\begin{flalign}
	\chi_i(\tau) &= -B_\varphi^\transpose \Phi_\varphi^\transpose(\tau^+,\tau)W_\varphi^{-1}(\tau^+, T-\tau^-)\xi_i(\tau^+), \hspace{-1cm} &
	\label{eq:chi}
\end{flalign}
in which 
\begin{align}
	W_\varphi(\tau^+, \tau) &= \INT*{\tau^+}{\tau}{ \Phi_\varphi(\tau^+, \bar{\tau}) B_\varphi B_\varphi^\transpose \Phi_\varphi^\transpose(\tau^+, \bar{\tau}) }{\bar{\tau}}
\end{align}
is the \emph{controllability Grammian} (see, e.g., \cite[Eq. (5-13, 5-14)]{Chen1984}). The inverse of the latter exists if \eqref{eq:xi:ode} is controllable. Hence, it remains to verify this property for \eqref{eq:xi:ode}. Obviously, $\varphi_{d,i}(\tau)$ is a flat output for \eqref{eq:xi:ode} by \eqref{eq:diffParam:m_n} and \eqref{eq:xi:state} implying the controllability of \eqref{eq:xi:ode} (see, e.g., \cite[Sec. 3.2.2]{Sira-Ramirez2004}). With this, inserting \eqref{eq:chi} in the solution
\begin{align}
	\xi_i(\tau) &= \Phi_\varphi(\tau,\tau^+)\big(\xi_i(\tau^+) + \INT*{\tau^+}{\tau}{\Phi_\varphi(\tau^+, \bar{\tau}) B_\varphi \chi_i(\bar{\tau}) }{\bar{\tau}}
	\big),
\end{align}
of \eqref{eq:xi:ode} on $\Time_2$ the result
\begin{align}
	\xi_i(\tau) &= \Phi_\varphi(\tau, \tau^+)\big( I_{n_\eta n_\n}
	\notag\\&\quad - W_\varphi(\tau^+, \tau)W_\varphi^{-1}(\tau^+, T-\tau^-) \big) \xi_i(\tau^+)
	\label{eq:xi:solution:2}
\end{align}
for $\tau\in\Time_2$ is obtained. Consequently, a constructive approach for the posed controllability problem has been found, which also yields a solution of the kernel equations \eqref{eq:faultDiagnosisKernel}. Furthermore, it leads to the easy verifiable condition \eqref{eq:identifiability}. The results of this section are summarized in the following theorem.
\begin{thm}\label{thm:identifiability}\textbf{\emph{(Solution of the kernel equations)}}
	Assume $T>T_0$ and that \eqref{eq:identifiability} holds for $i=1,\dots,n_f$. Then, the fault diagnosis kernel equations \eqref{eq:faultDiagnosisKernel} have a piecewise $C$-solution.
\end{thm}
If the condition of Theorem \ref{thm:identifiability} is fulfilled, then 
the fault $f_i(t)$ is identifiable according to Theorem \ref{thm:identification}. The condition \eqref{eq:identifiability} only depends on system parameters, and can thus be checked a priori. 

Note that in the case $\bar{d}(t)\not\equiv 0$, Theorem \ref{thm:identifiability} implies the ability to detect and estimate a fault $f_i(t)$ by making use of the results in Theorem \ref{thm:detection}.

\section{Example}\label{sec:example}
The application of the presented fault diagnosis  approach is illustrated for a faulty $4\times4$ heterodirectional ODE-PDE system. The parameters of the considered system are $\Gamma(z) = \textstyle\diag*{z+1, 3 - \frac{z}{2}, \frac{z}{2} - 3, -\frac{1}{2}}$, 
\begin{align}
		A(z) &= 
		\begin{bmatrix*}
			 0 & 0 & a(z) & 0\\  
			 0 & 0 & a(z) & 0\\  
			 0 & a(z) & 0 & 0\\  
			 0 & a(z) & 0 & 0 
		\end{bmatrix*}
		\text{~with~}
		a(z) = \tfrac{1}{z-2},
		\\
		D(z,\zeta) &=
		\begin{bmatrix*} 
		0 & 0 & 0 & 0\\
		0 & -\tfrac{{\mathrm{e}}^{2\,z-2\,\zeta }}{10} & \tfrac{{\mathrm{e}}^{2\,z-2\,\zeta }}{10} & 0\\
		0 & \tfrac{{\mathrm{e}}^{2\,z-2\,\zeta }}{10} & -\tfrac{{\mathrm{e}}^{2\,z-2\,\zeta }}{10} & 0\\
		0 & 0 & 0 & 0 
		\end{bmatrix*}.
\end{align}
The BC \eqref{eq:sys:bc0} is determined by
\begin{align}
	Q_0 &= \begin{bmatrix*} 0 & -1\\  -0.5 & 0 \end{bmatrix*},
	&
	H_2 &= \begin{bmatrix*} 1 & 1\\  0 & 0 \end{bmatrix*}
\end{align}
and the BC \eqref{eq:sys:bc1} by
\begin{align}
	Q_1 &= \begin{bmatrix*} 0 & 0.5\\  1 & 0 \end{bmatrix*}
\end{align}
 as well as $B_3 = I_2$. For the ODE subsystem 
\begin{align}
	F &= \begin{bmatrix*} 0 & 1\\  -26 & -10 \end{bmatrix*}
\end{align}
and $L_2 = e_{2,2} e_{2,4}^\transpose$ are obtained. The disturbance inputs are characterized by $G_1(z) = \textstyle e_{1,4} e_{1,3}^\transpose (\bar{\theta}(z-\tfrac{1}{5})-\bar{\theta}(z-\tfrac{4}{5}))$ with $\bar{\theta}(z)$ as the step function and $G_3 = \vect{0 & I_2}$. Its deterministic part is $\tilde{d}(t) = d_1 \sin(0.5t + \alpha_d)$, $t\geq0$, with unknown amplitude $d_1\in\R$, phase $\alpha_d\in\R$ and $\tilde{G} = e_{1,3}$. A possible choice for the signal model is
\begin{align}
	S_d &= \MATRIX{0 & -0.5 \\ 0.5 & 0},
	&
	\tilde{R}_d &= \MATRIX{0 & 1}
\end{align}
with unknown IC $v_d(0)$. The bounded part $\bar{d}(t)$ of the disturbance is specified by $\bar{G} = \col{0 & I_2}$ and has the known bound $\delta = \col{0.7 & 0.3}$. All remaining system parameters are zero.

In order to demonstrate the generality of the approach, the considered system is subject to a component fault $f_1(t)$ at the boundary $z=0$ specified by $E_2 = e_{1,2} e_{1,3}^\transpose$, an actuator fault $f_2(t)$ affecting the input $u_2(t)$ at the boundary $z=1$ with $E_3 = e_{2,2} e_{2,3}^\transpose$  and a sensor fault $f_3(t)$ acting on the output $y_1(t)$ described by $E_5 = e_{1,2} e_{3,3}^\transpose$. The faults $f_i(t)$ are assumed to occur at unknown time instants $t_i>0$, which implies $f_i(t) = 0$ for $t<t_i$. Hence, the corresponding signal models are defined for $t\geq t_i$. 
The fault $f_1(t) = f^0_1 \sin(\frac{1}{3}(t-t_1) + \alpha_f)$, $t\geq t_1$ with the unknown parameters $f^0_1, \alpha_f\in\R$, has the signal model
\begin{align}
	S_{f1} &= \MATRIX{0 & -\frac{1}{3} \\ \frac{1}{3} & 0},
	&
	\tilde{r}_{f1}^\transpose &= \MATRIX{0 & 1}.
\end{align}
Therein, the unknown IC $v_f^1(t_1) \not = 0$ specifies $f_1(t)$, $t\geq t_1$. The ramplike fault $f_2(t) = (f_2^0 + f_2^1 (t - t_2))$ for $t \geq t_2$ with $f_2^0, f_2^1\in \R$ is described by the signal model
\begin{align}
	S_{f2} &= \MATRIX{0 & 1 \\ 0 & 0}, &
	\tilde{r}_{f2}^\transpose &= \MATRIX{1 & 0}
\end{align}
with unknown IC $v_f^2(t_2)\not= 0$ for $t\geq t_2$. The fault $f_3(t) = f^0_3$, $t\geq t_3$ with the unknown amplitude $f^0_3\in\R$ yields the signal model described by $S_{f3} = 0$ and $\tilde{r}_{f3}=1$ with IC $v_f^3(t_3)\not= 0$ for $t\geq t_3$. The matrices of the common signal model for all faults are $S_f = \diag*{S_{f1}, S_{f2}, S_{f3}}$ and $\tilde{R}_f = \diag*{\tilde{r}_{f1}^\transpose, \tilde{r}_{f2}^\transpose, \tilde{r}_{f3}}$. 
Note that \eqref{eq:identifiability} is satisfied for this system, although there are only two measurements $y(t)\in\R[2]$ available. Hence, all three faults are identifiable (see Theorem \ref{thm:identifiability}) and decoupled from $\tilde{d}(t)$. 

\begin{figure} 
	\begin{tikzexternal}[threshold][true]\begin{tikzpicture}[baseline, trim axis right]
		\begin{axis}[
		name = main,
		ylabel = {$f_{B,i}(T)$}, 
		xlabel = {$T$},
		scale only axis,
		width = 7*0.7cm,
		height = 2cm,
		]
		\pgfplotstableread{data/thresholds.dat}\loadedtable
		\addplot [black, solid] table[x=T, y = fb1]{\loadedtable};
		\label{plt:fb1}
		\addplot [blue, dotted] table[x=T, y = fb2]{\loadedtable};
		\label{plt:fb2}
		\addplot [red, densely dashed] table[x=T, y = fb3]{\loadedtable};
		\label{plt:fb3}
		\end{axis};
		\begin{axis}[
		ylabel = {$f_{B,i}(T)$}, 
		xmin = 30,
		xmax = 50,
		clip = true,
		scale only axis,
		width = 4*0.7cm,
		height = 1.3cm,
		axis background/.style={fill=white},
		at = (main.north east),
		anchor = north east,
		xlabel = {},
		name = lens,
		legend style={
			draw = none,
			legend style={fill=none},
			cells={anchor=east},
			legend pos=outer north east,
		},
		]
		\pgfplotstableread{data/thresholds.dat}\loadedtable
		\addplot [black, solid] table[x=T, y = fb1]{\loadedtable};
		\addlegendentry{$f_{B,1}$} 
		\addplot [blue, dotted] table[x=T, y = fb2]{\loadedtable};
		\addlegendentry{$f_{B,2}$}
		\addplot [red, densely dashed] table[x=T, y = fb3]{\loadedtable};
		\addlegendentry{$f_{B,3}$}
		\end{axis};
		\end{tikzpicture}\end{tikzexternal}
	\caption{Thresholds $f_{B,i}$, $i=1,2,3$, in dependence of the detection time $T$.}\label{fig:thresholds}
\end{figure}
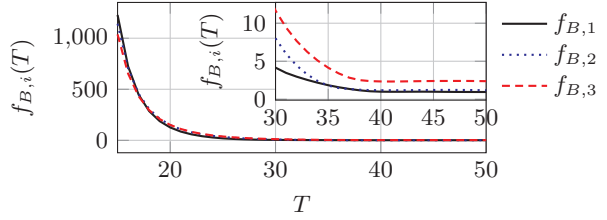

For solving the fault diagnosis kernel equations \eqref{eq:kernel:faultdiagnosis:MP}, \eqref{eq:requirement:Q:IE} and \eqref{eq:requirement:Q:ode}, firstly the kernel \eqref{eq:kernelEquation:backstepping} of the backstepping transformation must be computed. It is computed by an implementation of the successive approximation described in \cite{Hu2019} using MATLAB. For this, $z$ and $\zeta$ are discretized with \val{backsteppingGrid} points. The successive approximation is stopped when the maximal absolute value of the next iteration step is less than {\sisetup{scientific-notation = true}\val*{approxError}}, which requires \val{successizeSteps} steps. 
Secondly, a suitable $T$ must be determined. From \eqref{eq:transportationTimes} and \eqref{eq:Tmin} the lower bound for the detection time follows as $T_0 = \val{T0}$. Then, $T>T_0$ can be chosen to obtain an acceptable threshold value $f_{B,i}$. To this end, $f_{B,i}$ is computed for several $T$. The result shown in Figure \ref{fig:thresholds} indicates that a compromise has to be made between fast identification time, i.e., a small $T$, and a small threshold value $f_{B,i}$.
For the following $T=\val{T}$ is chosen. 
Then, $\xi_i(\tau)$ can be computed with \eqref{eq:xi:solution:2}, yielding $\varphi_{d,i}(\tau)$ in view of \eqref{eq:xi:output}. With this, the kernels $\tilde{m}_i(z,\tau)$, $\tilde{n}_i(\tau)$ and $\eta_i(\tau)$ are computed with \eqref{eq:diffParam:eta} and \eqref{eq:diffParam:pde:phi}. For a visualization, the resulting $e_{1,2}^\transpose\tilde{m}_{d,1}^\n(0,\tau)$ is shown in Figure \ref{fig:mn0}.
Using $\delta$, the thresholds $f_{B,1} = \val*{fb1}$, $f_{B,2} = \val*{fb2}$ and $f_{B,3} = \val*{fb3}$ are obtained.
In order to implement the fault diagnosis equations, the time integrations in the terms of \eqref{eq:fTilde} have to be discretized. By using a compound trapezoidal rule (see e.g. \cite[Sec. 5.3.2]{Krommer1998}), they can be implemented as finite impulse response filters, operating on the sampled input and output. More details about the time discrete implementation can be found in \cite{Kiltz2014}, where similar expressions are used for the fault detection of nonlinear finite-dimensional systems. All following simulations and approximations of the integral terms are processed with a step size of {\sisetup{scientific-notation = true}\val*{dt}}.

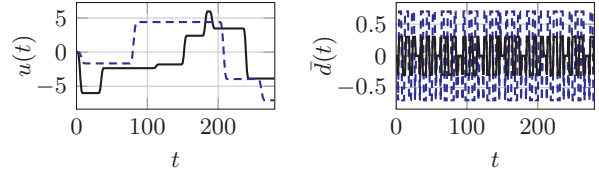
\begin{figure}
	\centering
	\pgfplotsset{every axis/.append style={
			width = 0.5\columnwidth,
			height=3cm,
		}
	}
	\begin{tikzexternal}[excitation][false]
		\pgfplotsset{every axis/.append style={	xlabel = {$t$},}}
		\begin{tikzpicture}[
		baseline, 
		trim axis right, 
		inner frame sep = -.5ex]
		\begin{axis}[
		name = u,
		ylabel = {$u(t)$},
		ylabel style={yshift = -.7em}
		]
		\pgfplotstableread{data/signals.dat}\loadedtable
		\addplot [black, solid] table[x = t, y = u2]{\loadedtable};
		\label{plot:u2}
		\addplot [blue, densely dashed] table[x = t, y = u1]{\loadedtable};
		\label{plot:u1}		
		\end{axis};
		\begin{axis}[
		anchor = left of south west, 
		at = (u.right of south east),
		xshift = 1em,
		name = dBounded,
		ylabel = {$\bar{d}(t)$},
		clip = true,
		ylabel style={yshift = -.7em}
		]
		\pgfplotstableread{data/signals.dat}\loadedtable
		\addplot [blue, densely  dashed] table[x=t, y = d1]{\loadedtable};
		\label{plot:d1}
		\addplot [black] table[x=t, y = d2]{\loadedtable};
		\label{plot:d2}
		\end{axis}
		\end{tikzpicture}
	\end{tikzexternal}
	\caption{Signals $u_1(t)$ \plotref{plot:u1}, $u_2(t)$ \plotref{plot:u2}, $\bar{d}_1(t)$ \plotref{plot:d1} and $\bar{d}_2(t)$ \plotref{plot:d2} used for the simulation.}
	\label{fig:simulation}
\end{figure}

For the simulation of the faulty system in MATLAB, a finite-dimensional model is used that results from the finite difference method. It is simulated with vanishing ICs, the signals for $f(t)$ and $\tilde{d}(t)$, specified before as well as $u(t)$ shown in Figure \ref{fig:simulation}. The faults occur at $t_1 = \val{occurrence1}$, $t_2 = \val{occurrence2}$ and $t_3 = \val{occurrence3}$ so that the assumption $T_0 < \Delta t < t_{i+1}-t_i$ (see Section \ref{sec:problem}) holds. 

\begin{figure}
	\pgfplotsset{every axis/.append style={
			height=3cm,
			xlabel style={yshift=.5em},
			extra x tick style = {
				tick label style = {anchor = south east, xshift=.66mm},
				tick align = center,
				tick style = {thick, black},
			},
		}	
	}	
	\centering
	\begin{tikzexternal}[identification][true]
		\begin{tikzpicture}[line join=round, remember picture, trim axis right] 
		\begin{axis}[
		name = fI, 
		ylabel={$f_1(t)$, $\hat{f}_1(t)$},
		extra x ticks = {\foI},
		extra x tick labels = {$t_{1}$},
		]
		\markDomains
		\addplot [red, solid]  table[x=t, y = f1]{data/identification.dat};	\label{plot:identification:f1:hat}
		\addplot [black, dashed] table[x=t, y = f1]{data/signals.dat};	\label{plot:identification:f1:sim}
		\end{axis}

		\begin{axis}[
		name = fII, 
		anchor = above north west,
		at={(fI.below south west)}, 
		ylabel={$f_2(t)$, $\hat{f}_2(t)$},
		extra x ticks = {\foII},
		extra x tick labels = {$t_2$},
		]
		\markDomains
		\addplot [red, solid]  table[x=t, y = f2]{data/identification.dat};	\label{plot:identification:f2:hat}
		\addplot [black, dashed] table[x=t, y = f2]{data/signals.dat};	\label{plot:identification:f2:sim}
		\end{axis}
		
		\begin{axis}[
		name = fIII, 
		anchor = above north west,
		at={(fII.below south west)}, 
		ylabel={$f_3(t)$, $\hat{f}_3(t)$},
		ymin = -50,
		extra x ticks = {\foIII},
		extra x tick labels = {$t_3$},
		]
		\markDomains
		\addplot [red, solid]  table[x=t, y = f3]{data/identification.dat};	\label{plot:identification:f3:hat}
		\addplot [black, dashed] table[x=t, y = f3]{data/signals.dat};	\label{plot:identification:f3:sim}
		\end{axis}
		\end{tikzpicture}
	\end{tikzexternal}
	\caption{Fault identification results $\hat{f}_{i}(t)$, $i=1,2,3$ \plotref{plot:identification:f2:hat} for the fault $f_i(t)$ \plotref{plot:identification:f2:sim} with the detection window of length $T$ marked by \noIdentificationAreaLabel assuming $\bar{d}\equiv 0$.}\label{fig:identification}
\end{figure}
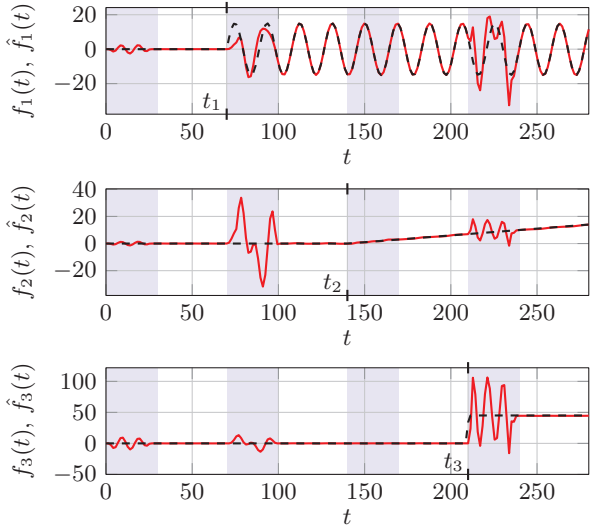

\paragraph*{Fault identification.} At first, the fault identification problem is solved for the considered setup, i.e., the system is simulated with $\bar{d} \equiv 0$. The identified faults $\hat{f}_i(t)$ are presented in Figure \ref{fig:identification}.
All faults are identified in finite time given by the detection window length $T$. For $0\leq t < T$ the estimation errors do not vanish, because the disturbance $\tilde{d}(t)$ has to be determined within the detection window. Then, the fault estimation errors vanishes until a fault occurs. Obviously, the fault estimates are coupled within the detection window, i.e., the fault $f_1(t)$ at $t_1$ leads to a non vanishing estimate $\hat{f}_2$. However, after $t>t_1+T$ it can be verified that no fault $f_2$ and $f_3$ has occurred. The same result also holds for the faults occurring at $t_2$ and $t_3$.

\begin{figure}
	\pgfplotsset{every axis/.append style={
			height=3cm,
			xlabel style={yshift=.5em},
		}
	}
	\centering

	\begin{tikzexternal}[detection]
		\begin{tikzpicture}[line join=round, remember picture, trim axis right] 
		\begin{axis}[
		name = fI, 
		ylabel={$f_1(t)$, $\hat{f}_1(t)$},
		extra x ticks = {\tTDI},
		extra x tick style = {
			grid style = {dashed, black, semithick},
			tick label style = {anchor = south east},
			tick align = center,
			tick style = {thick, black},		
		},
		extra x tick labels = {$\hat{t}_{1}$},
		]
		\markDomains;
		\pgfplotstableread{data/detection1.dat}{\detectionTableI};
		\addplot [red, solid]  table[x=t, y = f_hat]{\detectionTableI};
		\label{plot:detection:f1:hat}
		\addplot [black, dashed] table[x=t, y = f1]{data/signals.dat};
		\label{plot:detection:f1:sim}
		\addplot [blue, densely dotted, restrict x to domain  = \fT:\foIT]  table[x=t, y = bound_lower]{\detectionTableI};
		\addplot [blue, densely dotted, restrict x to domain  = \fT:\foIT]  table[x=t, y = bound_upper]{\detectionTableI}; \label{plot:detection:fB1}
		\addplot [green, dashdotted, restrict x to domain  = \foIT:\tEnd]  table[x=t, y = f_hat_bound_upper]{\detectionTableI};
		\addplot [green, dashdotted, restrict x to domain  = \foIT:\tEnd]  table[x=t, y = f_hat_bound_lower]{\detectionTableI};
		\label{plot:detection:f1:hat:bound}		
		\end{axis}
		
		\begin{axis}[
		name = fII, 
		anchor = above north west,
		at={(fI.below south west)}, 
		ylabel={$f_2(t)$, $\hat{f}_2(t)$},
		extra x ticks = {\tTDII},
		extra x tick style = {
			grid style = {dashed, black, semithick},
			tick label style = {anchor = south east},
			tick align = center,
			tick style = {thick, black},		
		},
		extra x tick labels = {$\hat{t}_{2}$},		
		]
		\markDomains;
		\pgfplotstableread{data/detection2.dat}{\detectionTableII};
		\addplot [red, solid]  table[x=t, y = f_hat]{\detectionTableII};
		\label{plot:detection:f2:hat}
		\addplot [black, dashed] table[x=t, y = f2]{data/signals.dat};
		\label{plot:detection:f2:sim}
		\addplot [blue, densely dotted, restrict x to domain  = \fT:\tTDII]  table[x=t, y = bound_lower]{\detectionTableII};
		\addplot [blue, densely dotted, restrict x to domain  = \fT:\tTDII]  table[x=t, y = bound_upper]{\detectionTableII}; \label{plot:detection:fB2}
		\addplot [green, dashdotted, restrict x to domain  = \tTDII:\tEnd]  table[x=t, y = f_hat_bound_upper]{\detectionTableII};
		\addplot [green, dashdotted, restrict x to domain  = \tTDII:\tEnd]  table[x=t, y = f_hat_bound_lower]{\detectionTableII};
		\label{plot:detection:f2:hat:bound}		
		\end{axis}
		
		\begin{axis}[
		name = fIII, 
		anchor = above north west,
		at={(fII.below south west)}, 
		ylabel={$f_3(t)$, $\hat{f}_3(t)$},
		extra x ticks = {\tTDIII},
		extra x tick style = {
			grid style = {dashed, black, semithick},
			tick label style = {anchor = south west},
			tick align = center,
			tick style = {thick, black},		
		},
		extra x tick labels = {$\hat{t}_{3}$},		
		]
		\markDomains;
		\pgfplotstableread{data/detection3.dat}{\detectionTableIII};
		\addplot [red, solid]  table[x=t, y = f_hat]{\detectionTableIII};
		\label{plot:detection:f3:hat}
		\addplot [black, dashed] table[x=t, y = f3]{data/signals.dat};
		\label{plot:detection:f3:sim}
		\addplot [blue, densely dotted, restrict x to domain  = \fT:\foIIIT]  table[x=t, y = bound_lower]{\detectionTableIII};
		\addplot [blue, densely dotted, restrict x to domain  = \fT:\foIIIT]  table[x=t, y = bound_upper]{\detectionTableIII}; \label{plot:detection:fB3}
		\addplot [green, dashdotted, restrict x to domain  = \foIIIT:\tEnd]  table[x=t, y = f_hat_bound_upper]{\detectionTableIII};
		\addplot [green, dashdotted, restrict x to domain  = \foIIIT:\tEnd]  table[x=t, y = f_hat_bound_lower]{\detectionTableIII};
		\label{plot:detection:f3:hat:bound}		
		\end{axis}		
		\end{tikzpicture}
	\end{tikzexternal}
	\caption{Fault detection and estimation results $\hat{f}_{i}(t)$, $i=1,2,3$ \plotref{plot:identification:f2:hat} for the fault $f_i(t)$ \plotref{plot:identification:f2:sim} with the detection window of length $T$ marked by \noIdentificationAreaLabel, the thresholds $f_{B,i}$ \plotref{plot:detection:fB3} for the fault detection and the bounds of the fault estimation error $\hat{f}_i (t) \pm f_{B,i}$   \plotref{plot:detection:f3:hat:bound} in the presence of $\bar{d}$.}\label{fig:detection}
\end{figure}
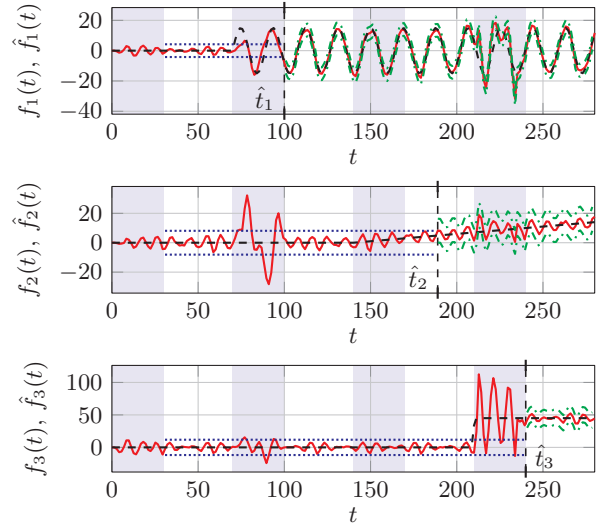

\paragraph*{Fault detection.} In a second simulation the influence of a bounded disturbance is investigated. Hence, only fault detection and estimation is possible. 
The bounded disturbance is chosen so that the worst case $\abs{f_i(t) - \tilde{f}_i(t)} = f_{B,i}$, $i=1,2,3$, is assumed for some $t$. This yields $\bar{d}_i(t)$ as shown in Figure \ref{fig:simulation}. With this disturbance $\bar{d}(t)$ and the signals used before, the second simulation is performed (see Figure \ref{fig:detection}). After the corresponding threshold is exceeded for $t>t_i+T$, the fault $f_1(t)$ and $f_3(t)$ are detected at $\hat{t}_i = t_i+T$, $i=1,3$, and $f_2(t)$ at $\hat{t}_3 = \val*{tDetectionT2}$. Note that $f_2(t)$ is a sliding fault with a small slope, which is in general difficult to detect.
The fault estimates are bounded by $f_{B,i}$ in finite time $T$, after the beginning $t_0=0$ and the fault occurrences $t_i$, $i=1,2,3$. It can be seen that the detection delay, i.e., $t_{\Delta}^i = \hat{t}_i - t_i$, is dependent on $T$ and $f_{B,i}$. On the one hand, $t_{\Delta}^i$ is lower bounded by $T$, which is the case for $\hat{t}_i$, $i=1,3$. Hence, a small $T$ could lead to a faster detection. On the other hand, a small $f_{B,i}$ (i.e., $T$ larger, see Figure \ref{fig:thresholds}), could lead to an earlier detection of $f_2(t)$. Furthermore, a lower threshold value allows more accurate estimates, since the fault estimation error is reduced. 

\section{Concluding remarks}
The simulation results suggest that the detection delay may be considerably reduced with a more detailed investigation of the intervals $t\in[t_i, t_i+T)$, since the thresholds $f_{B,i}$, $i=1,3$, are already exceeded inside of these intervals. This requires to take the transient behaviour of $\hat{f}$ for the fault detection into account, which is an interesting topic for further research.

Further interesting results could be achieved by considering stochastic disturbances and by extending the approach to PDEs with higher-dimensional spatial domain. 


\appendix

\section{Determination of the input-output equation \textbf{(\ref{eq:io})}}\label{sec:ioEquation}
Apply \eqref{eq:transformation:M} to \eqref{eq:sys:pde} resulting in
\begin{align}
&\innerp{M& x'(t)}_{\Omega,\Time} 
= \innerp{M& \Gamma\dt x(t)}_{\Omega,\Time}
+ \innerp{M& A x(t)}_{\Omega,\Time}
\notag\\&\quad
+ \innerp{M & A_0 x^{\n}(0,t)}_{\Omega,\Time}
+ \innerp{M & \mathcal{D}[x(t)]}_{\Omega,\Time}
\notag\\&\quad
+ \innerp{M& H_1 w(t)}_{\Omega,\Time}
+ \innerp{M& B_1 u(t)}_{\Omega,\Time}
\notag\\&\quad
+ \innerp{M& E_1 f(t)}_{\Omega,\Time}
+ \innerp{M& G_1 d(t)}_{\Omega,\Time}
\label{eq:derivation:M:1}\end{align}
with $\mathcal{D}[x(t)](z) = \INT{0}{z}{D(z,\zeta)x(\zeta,t)}{\zeta}$. In order to eliminate the terms in \eqref{eq:derivation:M:1} depending on $x(z,t)$, integration by parts \wrt $z$ for the left-hand-side in \eqref{eq:transformation:M} is applied. This yields
\begin{align}\label{eq:integrationByParts:z}
	\innerp{M& x'(t)}_{\Omega,\Time} 
	&= \innerp{ M(1) & x(1,t)}_{\Time} - \innerp{ M(0) & x(0,t)}_{\Time}
	\notag\\&\quad -\innerp{ M' & x(t)}_{\Omega,\Time}.
\end{align}
Similarly, for an integration by parts \wrt $\tau$ for the first term in the right-hand-side of \eqref{eq:derivation:M:1}, use $\dt x(z,t-\tau) = -\dtau x(z,t-\tau)$. This leads to
\begin{flalign}
\innerp{ M& \Gamma \dtau x(t)}_{\Omega,\Time} 
	&= \innerp{\Gamma^\transpose M(T) & x(t-T)}_\Omega 
	\notag\\&\hspace*{-4em} - \innerp{\Gamma^\transpose M(0) & x(t)}_\Omega
	- \innerp{ \Gamma^\transpose \dtau M & \Gamma x(t)}_{\Omega,\Time}.
	\label{eq:integraionByParts:tau}\hspace*{-3em}&\end{flalign}
Consequently, the unknown states $x(z,t)$ and $x(z,t-T)$ are eliminated by imposing \eqref{eq:requirement:M:IE}. For the integral term in \eqref{eq:derivation:M:1}, changing the order of integration, i.e.,
\begin{align}
&\INT*{0}{1}{ \INT*{0}{z}{ M^\transpose(z,\tau) D(z, \zeta) x(\zeta, \tau) }{\zeta} }{z}
\notag\\
&=
\INT*{0}{1}{\left(\INT*{z}{1}{ D^\transpose(\zeta, z) M(\zeta,\tau) }{\zeta}\right)^\transpose \, x(z,\tau) }{z}
\end{align}
leads to
\begin{align}
\innerp{M & \mathcal{D}[x]}_{\Omega,\Time} = \innerp{\mathcal{D}^*[M] & x}_{\Omega,\Time},
\label{eq:adjoint:term:D}
\end{align}
where $\mathcal{D}^*$ (see \eqref{eq:Dstar}) is the formal adjoint of $\mathcal{D}$. Then, using \eqref{eq:integrationByParts:z}, \eqref{eq:integraionByParts:tau} with \eqref{eq:requirement:M:IE}, \eqref{eq:adjoint:term:D} in \eqref{eq:derivation:M:1} and straightforward manipulations to shift the remaining matrices to the other arguments yields
\begin{flalign}
&\innerp{- M' - \Gamma \dot{M} - A^\transpose M - \mathcal{D}^*[M] & x(t)}_{\Omega,\Time} 
\notag\\
&= \innerp{ M(0) & x(0,t)}_{\Time} - \innerp{ M(1) & x(1,t)}_{\Time}
\notag\\&\quad
+ \innerp{\innerp{ A_0 & M }_\Omega & x^{\n}(0,t)}_{\Time}
+ \innerp{\innerp{ H_1 & M }_\Omega & w(t)}_{\Omega,\Time}
\notag\\&\quad
+ \innerp{\innerp{ B_1 & M}_\Omega & u(t)}_{\Omega,\Time}
+ \innerp{\innerp{ E_1 & M}_\Omega & f(t)}_{\Omega,\Time}
\notag\\&\quad
+ \innerp{\innerp{ G_1 & M}_\Omega & d(t)}_{\Omega,\Time}.
\label{eq:derivation:M:2}
\end{flalign}
Hence, \eqref{eq:derivation:M:2} becomes independent of $x(z,t)$ if \eqref{eq:requirement:pde} holds. The BCs \eqref{eq:sys:bc0} and \eqref{eq:sys:bc1} can be utilized to simplify $x(0,t)$ and $x(1,t)$ in \eqref{eq:derivation:M:2}. In view of \eqref{eq:states:x} and the BCs \eqref{eq:sys:bc0} as well as \eqref{eq:sys:bc1} the result
\begin{subequations}\begin{align}
	x(0,t) &= J_\p^\transpose \big(Q_0 x^{\n}(0, t) + H_2 w(t) + B_2 u(t)
	\notag\\&\quad
	+ E_2 f(t) + G_2 d(t) \big) + J_\n^\transpose x^{\n}(0, t)
	\\
	x(1,t) &= J_\p^\transpose x^{\p}(1,t) + J_\n^\transpose \big(Q_1 x^{\p}(1,t) + B_3u(t)
	\notag\\&\quad\
	+ E_3 f(t) + G_3 d(t)\big)
\end{align}\label{eq:boundaryValues}\end{subequations}
is obtained. Consequently, the terms with $x(0,t)$ and $x(1,t)$ in \eqref{eq:integrationByParts:z} become
\begin{subequations}\begin{align}
	&\innerp{ M(0) & x(0,t)}_{\Time} 
	= 
	\innerp{\left(Q_0^\transpose J_\p + J_\n\right) M(0) & x^{\n}(0,t)}_{\Time}
	\notag\\&
	+ \innerp{H_2^\transpose J_\p M(0) & w(t)}_{\Time}
	+ \innerp{B_2^\transpose J_\p M(0) & u(t)}_{\Time}
	\notag\\&
	+ \innerp{E_2^\transpose J_\p M(0) & f(t)}_{\Time}
	+ \innerp{G_2^\transpose J_\p M(0) & d(t)}_{\Time}
	\hspace{-2em}
\end{align}
	as well as
\begin{align}
	&\innerp{ M(1) & x(1,t)}_{\Time}
	= 
	\innerp{ \left(Q_1^\transpose J_\n + J_\p\right) M(1) & x^{\p}(1,t) }_{\Time}
	\notag\hspace*{-2em}\\&
	+ \innerp{ B_3^\transpose J_\n M(1) & u(t)}_{\Time}
	\notag\\&
	+ \innerp{ E_3^\transpose J_\n M(1) & f(t)}_{\Time}
	+ \innerp{ G_3^\transpose J_\n M(1) & d(t)}_{\Time}.
	\end{align}\label{eq:boundaryValues:transformed}\end{subequations}
Hence, 
\begin{flalign}\label{eq:derivation:M:3}
0 &= 
\innerp{\innerp{ A_0 & M }_\Omega + \left(Q_0^\transpose J_\p + J_\n \right) M(0) & x^{\n}(0, t)}_{\Time}
\notag\\&\quad
- \innerp{ \left( Q_1^\transpose J_\n + J_\p\right) M(1) & x^{\p}(1,t) }_{\Time}
\notag\\ &\quad
+ \innerp{H_2^\transpose M_\p(0) + \innerp{H_1 & M}_{\Omega} & w(t)}_{\Omega,\Time}
\notag\\&\quad
+ \innerp{\innerp{B_1 & M}_\Omega + B_2^\transpose M_\p(0) - B_3^\transpose M_\n(1) & u(t)}_{\Time}
\notag\\&\quad 
+ \innerp{\innerp{E_1 & M}_{\Omega} + E_2^\transpose M_\p(0) - E_3^\transpose M_\n(1) & f(t)}_{\Time}
\notag\\&\quad 
+ \innerp{\innerp{G_1 & M}_{\Omega} + G_2^\transpose M_\p(0) - G_3^\transpose M_\n(1) & d(t)}_{\Time}
&\end{flalign}
results from taking \eqref{eq:requirement:pde} and \eqref{eq:boundaryValues:transformed} into account.

The result \eqref{eq:derivation:M:3} still depends on the lumped state $w(t)$. In order to eliminate it, apply the transformation \eqref{eq:transformation:P} to \eqref{eq:sys:ode} and use the substitution $\dt w(t-\tau) = -\dtau w(t-\tau)$. This yields
\begin{align}\label{eq:derivation:P:1}
- \innerp{P & \dtau w(t)}_{\Time} 
&= \innerp{F^\transpose P & w(t)}_{\Time} 
+ \innerp{L_2^\transpose P & x^{\n}(0,t)}_{\Time}
\notag\\&\quad
+ \innerp{B_4^\transpose P & u(t)}_{\Time}
+ \innerp{E_4^\transpose P & f(t)}_{\Time}
\notag\\&\quad
+ \innerp{G_4^\transpose P & d(t)}_{\Time}.
\end{align}
For the left-hand side in \eqref{eq:derivation:P:1}, integration by parts leads to
\begin{align}
\innerp{P & \dtau w(t)}_{\Time} = -\innerp{\dtau P & w(t)}_{\Time}
\label{eq:integraionByParts:P}
\end{align}
if $P(\tau)$ satisfies \eqref{eq:requirement:P:IE}. Then, taking \eqref{eq:integraionByParts:P} in \eqref{eq:derivation:P:1} into account, 
\begin{align}
&\innerp{\dtau P - F^\transpose P & w(t)}_{\Time}
= 
\innerp{L_2^\transpose P & x^{\n}(0,t)}_{\Time}
\notag\\&
+ \innerp{B_4^\transpose P & u(t)}_{\Time}
+ \innerp{E_4^\transpose P & f(t)}_{\Time}
+ \innerp{G_4^\transpose P & d(t)}_{\Time}
\label{eq:derivation:P:2}\end{align}
results. Hence, $w(t)$ can be replaced by $u(t)$, $f(t)$ and $d(t)$ if \eqref{eq:requirement:ode:P} holds (see third line in \eqref{eq:derivation:M:3}). This leads to
\begin{flalign}\label{eq:derivation:M:4}
&0 = \innerp{N & x^{\n}(0, t)}_{\Time}
\\&
- \innerp{ \left( Q_1^\transpose J_\n + J_\p\right) M(1) & x^{\p}(1,t) }_{\Time}
+ \innerp{M_B & u(t)}_{\Time} \hspace{-.5em}
\notag\\&
+ \innerp{\innerp{E_1 & M}_{\Omega} + E_2^\transpose M_\p(0) - E_3^\transpose M_\n(1) + E_4^\transpose P& f(t)}_{\Time}\hspace*{-5em}
\notag\\&
+ \innerp{\innerp{G_1 & M}_{\Omega} + G_2^\transpose M_\p(0) - G_3^\transpose M_\n(1) + G_4^\transpose P & d(t)}_{\Time},\hspace*{-5em}
\notag&
\end{flalign}
by consideration of \eqref{eq:N} and \eqref{eq:MB}. In \eqref{eq:derivation:M:4}, $x^{\n}(0,t)$ is replaced with $y(t)$, $f(t)$ and $d(t)$ in view of \eqref{eq:sys:y}, which leads to
\begin{align}
\innerp{N & x^{\n}(0,t)}_{\Time}
&=
\innerp{ N & y(t)}_{\Time} 
- \innerp{E_5^\transpose N & f(t)}_{\Time}
\notag\\&\quad
- \innerp{G_5^\transpose N & d(t)}_{\Time}.
\label{eq:derivation:output}\end{align}
Taking \eqref{eq:derivation:output} in \eqref{eq:derivation:M:4} into account
\begin{align}
&0= 
\innerp{ N & y(t)}_{\Time}
- \innerp{ \left( Q_1^\transpose J_\n + J_\p\right) M(1) & x^{\p}(1,t) }_{\Time}
\notag\\&\quad
+ \innerp{M_B & u(t)}_{\Time}
+ \innerp{ M_E & f(t)}_{\Time}
+ \innerp{ M_G & d(t)}_{\Time}
\label{eq:derivation:M:5}
\end{align}
follows, in which $M_E(\tau)$ and $M_G(\tau)$ are defined in \eqref{eq:ME} and \eqref{eq:MG}.
Finally, the remaining unknown $x^{\p}(1,t)$ in \eqref{eq:derivation:M:4} is eliminated due to \eqref{eq:requirement:BC:1}. Then, \eqref{eq:io} is obtained in light of \eqref{eq:disturbance}.

\begin{ack}
This work was funded by the Deutsche Forschungsgemeinschaft (DFG, German Research Foundation) under individual grant reference 391022641.
\end{ack}

\bibliographystyle{plain}        
\bibliography{Fehlerdetektion}   

\end{document}